\documentclass[11pt]{article}
\usepackage{amsmath,amssymb, mcite}
\usepackage{graphicx}
\usepackage{multirow}
 \usepackage{tikz}
 \usetikzlibrary{calc}
 \usetikzlibrary{patterns}
 \usetikzlibrary{decorations.pathreplacing}
 \usetikzlibrary{decorations.markings}
 \usetikzlibrary{decorations.pathmorphing}
 \usetikzlibrary{positioning}

\def\beq{\begin{equation}}
\def\eeq{\end{equation}}
\def\bea{\begin{eqnarray}}
\def\eea{\end{eqnarray}}
\def\ba{\begin{array}}
\def\ea{\end{array}}
\def\bit{\begin{itemize}}
\def\eit{\end{itemize}}

\def\nn{\nonumber}

\def\al{\alpha}
\def\ga{\gamma}

\def\si{\sigma}

\def\a{\alpha}
\def\b{\beta}

\def\pq{(p \cdot q)}

\def\a{\alpha}

\def\b{\beta}

\def\cO{{\cal O}}

\def\cA{{\cal A}}

\begin{document}

\tikzset{
    photon/.style={decorate, decoration={snake}, draw=black},
    electron/.style={draw=black, postaction={decorate},
        decoration={markings,mark=at position .55 with {\arrow[draw=black]{>}}}},
        aelectron/.style={draw=black, postaction={decorate},
        decoration={markings,mark=at position .55 with {\arrow[draw=black]{<}}}},
    gluon/.style={decorate, draw=black,
        decoration={coil,amplitude=4pt, segment length=5pt}} 
}

\begin{center}
{\bf \Large A rederivation of the conformal anomaly for spin-$\frac12$}\\[1cm]
{\bf Hadi Godazgar and Hermann Nicolai}\\[0.5cm]
{{\it 
Max-Planck-Institut f\"ur Gravitationsphysik\\
(Albert-Einstein-Institut)\\
M\"uhlenberg 1, D-14476 Potsdam, Germany\\
}}

\vspace{1cm}

January 4, 2018
\end{center}

\vspace{1cm}

{\footnotesize We rederive the conformal anomaly for spin-$\frac12$ fermions by a genuine
Feynman graph calculation, which has not been available so far. Although our calculation
merely confirms a result that has been known for a long time, the derivation is new, and 
thus furnishes a method to investigate more complicated cases (in particular concerning
the significance of the quantum trace of the stress tensor in non-conformal theories) 
where there remain several outstanding and unresolved issues.
}

\vspace{0.3cm}

\section{Introduction}
Conformal anomalies have been studied for a long time, see 
\cite{CapDuff,DDI,Duff1,Tsao,Perry,FT,FT1,FT2,DS,BGVZ,ST,Tseytlin} for original references
and \cite{Parker,EH,BD,Duff2,Vass,Barvinsky} for reviews and further references. 
In four dimensions the gravitational part of the conformal anomaly takes the form
\beq\label{A}
\cA \,= \, a \, {\rm E}_4 \,+\, b \,\Box R \,+\,  c \, C^{\mu\nu\rho\sigma}C_{\mu\nu\rho\sigma} 
\eeq
where $C_{\mu\nu\rho\sigma}$ is the Weyl tensor and ${\rm E}_4$ the Euler invariant.
Unlike the first and last term the middle contribution can in principle be removed by a local
counterterm ($\sim R^2$), but we will keep it here for later purposes. These three terms 
are the only local expressions which satisfy the Wess-Zumino consistency condition,
while an $R^2$ contribution would require a non-local completion of the anomaly for the consistency 
condition to be obeyed.

In this paper we give a new derivation of the coefficients $a,b,c$ for spin-$\frac12$
(Majorana) fermions, by directly calculating 
\begin{equation}
 \cA = g^{\mu \nu} \langle T_{\mu \nu} \rangle
\end{equation}
up to second order in the metric fluctuations, thus extending the $\cO(h)$ calculation of Capper 
and Duff \cite{CapDuff}. We note that the $b$ and $c$ coefficients were originally determined
from the two-point correlator of stress tensors in \cite{CapDuff} because the two-point function is 
renormalised by the same counterterm as the 3-point function \cite{Duff1}, 
but this calculation does not yield the $a$ coefficient.  In this paper, by going to $\cO(h^2)$, 
we find all coefficients `in one go'; there is thus no need to distinguish between type
$A$ and $B$ anomalies \cite{DS}, as both appear on an equal footing.
Of course, the coefficients of the spin-$\frac12$ conformal anomaly have been 
known for a long time and have been determined by various different methods, 
via one-loop divergences and heat kernel expansions \cite{Tsao, Perry, Parker,EH,BD,Vass}, 
conformal higher spins \cite{Tseytlin}, path integral methods \cite{Endo, BvanN}, or by QFT in curved spacetime methods \cite{Dowker, Brown}. 
Curiously, however, to the best of our knowledge, this computation has  never been 
done {\em \`a la} Capper-Duff up to $\cO(h^2)$. In fact, a derivation closest in spirit 
to the present one is in recent work by Bonora {\em et al.} \cite{Bonora1,Bonora2},
where, however, only the simpler parity odd contribution (related to the Pontryagin invariant) 
was considered. Our rederivation is, in principle, a straightforward calculation, very much
like the standard textbook  derivation of the axial anomaly via triangle graphs, though
far more cumbersome in practice. Notably, and in contrast to several
other derivations, it does not rely on kinematic choices, such 
as special gauges for the external graviton $h_{\mu\nu}$, nor special values for external 
momenta, nor on-shell conditions. It thus also provides a toolkit for a similar 
`textbook calculation' of the (again known) $s=0, 1$ anomalies that still remains to 
be done in this way.

To be sure, we basically regard the present derivation  as just a `warm-up' exercise for 
investigating the conformal anomaly in {\em non}-conformal theories, in particular 
for $s= \frac32$ (that is, Poincar\'e supergravity) where there remain several open issues.
These concern for instance the occurrence (or not) of $R^2$ and/or non-local contributions 
to the anomaly; a full clarification of non-local terms will probably require 
the full machinery of scalar $n$-point integrals that we review and 
further develop in section~4 of this paper. The dependence of the $a$ and 
$c$ coefficients on the choice of gauge for the external gravity fluctuations 
that has been observed for $s\geq \frac32$ \cite{CD,CD1,FT,FT2,Tseytlin} is a very strange 
feature, as it would seem to indicate a breakdown of general covariance -- whereas a proper
definition of the conformal anomaly should result in a gauge  invariant answer also
for non-conformal theories (this question is relevant for the possible cancellation of 
the $c$ coefficients for $N\geq 5$ Poincar\'e supergravities \cite{MN1}). Another open 
issue is to see precisely {\em why} the result for spin-$\frac32$ comes out to be negative
(this is the only field that contributes with a negative $c$ coefficient, and is thus indispensable 
for any cancellation), a feature that is probably related to the absence of 
a gauge invariant stress tensor and a positive definite Hilbert space of states for spin-$\frac32$.

The organisation of this paper is as follows: In section \ref{sec:weyl}, we give the Weyl transformation of the curvature, Ricci tensor and scalar and review the Weyl transformation properties of the actions for scalar, Dirac, Maxwell and gravitino fields. In section \ref{sec:maj}, we consider the action for a massless Majorana field and the expectation value of the stress tensor for such a theory. We present the Feynman rules and calculate the expectation value at first order, subsection \ref{sub:1order}, and calculate the $\Box R$ anomaly. We then consider the expectation value of the stress tensor at second order, subsection \ref{sub:2order}, and show that it is also conserved. We review and develop methods for calculating scalar 3-point loop integrals in section \ref{sec:3pointidentities}, which are then used to find the trace anomaly at second order in the metric perturbation in section \ref{sec:anom}. We provide a list of the expansion of some relevant quantities under metric perturbations, appendix \ref{app:exp}; give the result of scalar 2-point integrals, appendix \ref{app:int} and list some useful gamma matrix and integral identites, \ref{app:id}, in the appendices. We also relegate some technical calculations to appendices \ref{app:subanti2pt} and \ref{app:Ttilde}. 

A final word on our conventions. Lest our multiple use of Greek indices may raise confusion 
let us state once and for all the convention that we will follow throughout this paper: contractions 
with the full metric $g_{\mu\nu}$ are always  fully covariant, whereas the {\em flat} metric 
$\eta_{\mu\nu}$ is to be used for all contractions involving the metric fluctuations 
$h_{\mu\nu}$ or any quantities appearing inside Feynman diagrams. For instance, when 
writing out a contraction like $g^{\mu\nu} T_{\mu\nu}$ in terms of the metric fluctuations
$h_{\mu\nu}$ the result will be an infinite series in terms of the latter where now all 
contractions are w.r.t. to the Minkowski metric $\eta_{\mu\nu}$. Where appropriate we will
also use flat (Lorentz) indices $a,b,...$ in the fully covariant context, whereas the distinction
between flat and curved indices becomes void in terms of the fluctuation expansion.

\vspace{1cm}

\section{Preliminaries}

In this section we summarize some general results concerning Weyl transformations
so as to make our presentation self-contained, and for reference in future work.

\subsection{Weyl transformations} \label{sec:weyl}

We collect a list of the transformations of some tensors under a Weyl transformation 
\begin{equation}
 g_{\mu \nu} \longrightarrow \Omega^2 \, g_{\mu \nu} = e^{2 \sigma } \, g_{\mu \nu} ,
\end{equation}
where all quantities depend on $x$.  The curvature tensor,
\begin{equation}
 R_{\mu \nu \rho \sigma} = C_{\mu \nu \rho \sigma} + \frac{2}{d-2} g_{\mu[\rho} \, R_{\sigma]\nu} - \frac{2}{d-2} g_{\nu[\rho} \, R_{\sigma]\mu} - \frac{2}{(d-1)(d-2)} g_{\mu [\rho } g_{\sigma] \nu} R,
\end{equation}
and its contractions 
transform as follows:
\begin{align}
 R^{\mu}{}_{\nu \rho \sigma} \longrightarrow & \, R^{\mu}{}_{\nu \rho \sigma} - 2 \,\delta^{\mu}_{[\rho} \, \nabla^{}_{\sigma]} \nabla_{\nu} \sigma + 2 \,g^{\mu \al}\, g_{\nu[\rho}\, \nabla_{\sigma]} \nabla_{\al} \sigma + \delta^{\mu}_{[\rho} \, \partial^{}_{\sigma]} \sigma \, \partial_{\nu} \sigma \notag\\[3pt]
 &  - 2 \, g^{\mu \al} \,g_{\nu[\rho}\, \partial_{\sigma]} \sigma \,  \partial_{\al} \sigma - 2 \, \delta^{\mu}_{[\rho}\, g^{}_{\sigma] \nu}\, g^{\al \b}\, \partial_{\al} \sigma \, \partial_{\b} \sigma.\\[5pt]
 R_{\mu \nu} \longrightarrow & \,R_{\mu \nu} - (d-2)\, \nabla_{\mu} \nabla_{\nu} \sigma + (d-2)\, \partial_{\mu} \sigma \, \partial_{\nu} \sigma - g_{\mu \nu} \, \Box \sigma - (d-2) \, g_{\mu \nu}\, g^{\rho \sigma}\, \partial_{\rho} \sigma \, \partial_{\sigma} \sigma, \\[3pt]
 R \longrightarrow & \, \Omega^{-2} \Big[ R - 2(d-1) \, \Box \sigma  - (d-1)(d-2) \, g^{\mu \nu}\, \partial_{\mu} \sigma \, \partial_{\nu} \sigma \Big] \, .
  \end{align}
The covariant derivative also transforms under a Weyl transformation. In particular, the Christoffel symbol transforms as 
\begin{equation}
 \Gamma^{\rho}_{\mu \nu} \longrightarrow \Gamma^{\rho}_{\mu \nu} + 2 \, \delta^{\rho}_{(\mu} \, \partial^{}_{\nu)} \sigma - g^{\rho \sigma} g_{\mu \nu} \, \partial_{\sigma} \sigma,
\end{equation}
while the spin connection transforms as
\begin{equation} \label{spincontrans}
 \omega_{\mu}{}^{a b} \longrightarrow  \omega_{\mu}{}^{a b} + 2 e_{\mu}{}^{[a} e_{\nu}{}^{b]} g^{\nu \rho} \partial_{\rho} \sigma.
\end{equation}

\subsection{Weyl invariant actions for spins $s\leq 1$}

Given the transformation property of the quadratic operator 
\begin{align}
 \sqrt{-g} \left( - \Box + \frac{d-2}{4(d-1)} R \right) &\longrightarrow \Omega^{d-2} \sqrt{-g} \left( - \Box + \frac{d-2}{4(d-1)}  R \right)- \frac{d-2}{2} \Omega^2 \Box\si \notag \\[4pt]
& \hspace{25mm} - (d-2) \Omega^{d-2} g^{\mu \nu} \left(\frac{d-2}{4} \partial_{\mu} \sigma \partial_{\nu}\si
  + \partial_{\mu} \sigma \partial_{\nu}  \right) 
\end{align}
this operator is Weyl covariant if it acts on a scalar $\phi$ of conformal weight $-\frac{d-2}{2}$,
\begin{equation}
  \phi \longrightarrow \Omega^{-\frac{d-2}{2}} \phi.
\end{equation}
Furthermore, it is then clear that
\begin{equation}
  \sqrt{-g} \phi \left( - \Box +  \frac{d-2}{4(d-1)} R \right) \phi 
\end{equation}
is Weyl invariant.

For a spinor $\chi$ of conformal weight $-\frac{d-1}{2}$, 
\begin{equation}
  \chi \longrightarrow \Omega^{-\frac{d-1}{2}} \chi,
\end{equation}
the Dirac Lagrangian 
\begin{equation}
  \overline{\chi} \gamma^{\mu} D_{\mu} \chi   \equiv
  \overline{\chi} \gamma^{\mu}  \left( \partial_\mu + \frac14 \omega_{\mu\,ab} \ga^{ab}\right)\chi,
\end{equation}
 is already Weyl-invariant by itself without any modification, and for any $d$. This can be seen using the 
 transformation of the spin connection, \eqref{spincontrans}, and noting that 
 \begin{equation}
  \gamma_{\mu} \gamma^{\mu \nu} = (d-1) \gamma^{\nu}.
 \end{equation}
In four dimensions, the invariance of the Yang-Mills action is anyhow clear because of 
the invariance of the factor $\sqrt{-g} g^{\mu\rho} g^{\nu\si}$ multiplying 
${\rm Tr}( F_{\mu\nu} F_{\rho\si})$ under Weyl transformations (where the vector field 
$A_\mu$ is assigned Weyl weight zero). 
In arbitrary dimensions the Yang-Mills action is not, however, Weyl-invariant.

For completeness and later applications let us also display the action of 
a Weyl transformation on the Rarita-Schwinger action, which is not invariant.
It transforms as
\begin{equation}
 \epsilon^{\mu \nu \rho\sigma}  \overline{\psi}_{\mu} \gamma_5 \gamma_{\nu} \nabla_{\rho} \psi_{\sigma} \longrightarrow \Omega^{-4} \epsilon^{\mu \nu \rho\sigma}  \overline{\psi}_{\mu} \gamma_5 \gamma_{\nu} \nabla_{\rho} \psi_{\sigma} -2 i g^{\mu \rho} \partial_{\rho} \sigma \overline{\psi}_{[\mu} \gamma^{\nu} \psi_{\nu]},
\end{equation}
where 
\begin{equation}
 \psi_{\mu} \longrightarrow \Omega^{-1/2} \psi_{\mu}.
\end{equation}
Hence we see that Weyl invariance is already broken at the classical level. Indeed,
it is known that for spin-$\frac32$ one needs an action cubic in derivatives for
conformal invariance.

\section{Majorana fermions} \label{sec:maj}
In this paper we will consider only spin-$\frac12$ fermions as they appear to provide the simplest
context in which to perform the analysis up to $\cO(h^2)$. Accordingly, we start with the Dirac 
action for a Majorana fermion~\footnote{Up to an overall factor of $\frac12$ this 
action is the same for Dirac and Majorana fermions. The action for a massless Majorana fermion is also classically the same as the action for a Weyl fermion up to a total derivative term. There are recent claims that they are different at the quantum level and that there is, in particular, an odd parity anomaly for Weyl fermions \cite{Bonora1, Bonora2}. We will not address this claim here, but we just note that there is 
no issue for Majorana fermions as (\ref{Majorana}) is real.}: 
\begin{equation}\label{Majorana}
S= \frac{i}{2} \int e \overline{\chi}  \gamma^{\mu} D_{\mu} \chi = S^{(0)} + S^{(1)} + S^{(2)} + \dots, 
\end{equation}
where $D_{\mu}$ is the $\omega$-covariant derivative and $S^{(k)}$ is the action at order $k$ 
in the metric fluctuation $h_{\mu\nu}$, from 
\beq
g_{\mu\nu} (x) = \eta_{\mu\nu} + h_{\mu\nu}(x).
\eeq
Using the expansions in section \ref{app:exp}, we find that, up to second order in $h$,
\begin{align} \label{actionexp}
  S^{(0)} &= \frac{i}{4} \int  \overline{\chi} \overleftrightarrow{\partial\!\!\!/} \chi, \nonumber  \\[3pt]
S^{(1)} &=  - \frac{i}{8} \int \left( h^{\mu \nu} \, \overline{\chi} \gamma_{\mu} \overleftrightarrow{\partial_{\nu}} \chi- h \, \overline{\chi} \overleftrightarrow{\partial\!\!\!/}\chi\right),  \nonumber \\[3pt]
S^{(2)} &= \frac{i}{32} \int \Big(3\, h^{\mu \rho} h_{\rho}{}^{\nu} \,  \overline{\chi} \gamma_{\mu} \overleftrightarrow{\partial_{\nu}} \chi  - 2\,  h h^{\mu \nu} \,  \overline{\chi} \gamma_{\mu} \overleftrightarrow{\partial_{\nu}} \chi - 2\,  h^{\mu \nu} h_{\mu \nu} \, \overline{\chi} \overleftrightarrow{\partial\!\!\!/}\chi \notag \\[3pt]
& \hspace{70mm}+ h^2 \overline{\chi} \overleftrightarrow{\partial\!\!\!/}\chi +  h^{\sigma}{}_{\mu} \partial_{\nu} h_{\rho \sigma} \, \overline{\chi} \gamma^{\mu \nu \rho} \chi \Big),
\end{align}
where $h \equiv \eta^{\mu\nu} h_{\mu\nu}$,
$ \overleftrightarrow{\partial_{\mu}} =  \overrightarrow{\partial_{\mu}} - \overleftarrow{\partial_{\mu}}$,
and where the left action of the differential operator is only on the fermion $\overline{\chi}$.
Also, we use lower case Latin letters for tangent space indices, we use Greek indices for tensors after perturbatively expanding the metric. In both cases the position of indices is raised/lowered with the Minkowski metric.   
Moreover, the fermionic stress tensor admits a similar expansion, 
\begin{equation}
T_{\mu \nu} = \frac{2}{e} g_{\mu \rho} g_{\nu \sigma} \frac{\delta S}{\delta g_{\rho \sigma}} = - \frac{i}{4}  \left(  \overline{\chi} \gamma_{(\mu} \overleftrightarrow{D_{\nu)}} \chi - g_{\mu \nu} \overline{\chi} \overleftrightarrow{D\!\!\!/} \chi \right) = T^{(0)}_{\mu \nu} + T^{(1)}_{\mu \nu} + \dots ,
\end{equation}
where to first order in $h$, 
\begin{align} \label{Texp1}
  T^{(0)}_{\mu \nu} &=  - \frac{i}{4} \left( \, \overline{\chi} \gamma_{(\mu} \overleftrightarrow{\partial_{\nu)}} \chi- \eta_{\mu \nu} \, \overline{\chi} \overleftrightarrow{\partial\!\!\!/}\chi\right)   \\[3pt]
  T^{(1)}_{\mu \nu} &= - \frac{i}{8} \Big( h_{\rho (a} \,  \overline{\chi} \gamma^{\rho} \overleftrightarrow{\partial_{\nu)}} \chi + \eta_{\mu \nu} h^{\rho \sigma} \,  \overline{\chi} \gamma_{\rho} \overleftrightarrow{\partial_{\sigma}} \chi - 2\,  h_{\mu \nu} \, \overline{\chi} \overleftrightarrow{\partial\!\!\!/}\chi - \partial_{\rho} h_{\sigma (\mu} \, \overline{\chi} \gamma_{\nu)}{}^{\rho \sigma} \chi \Big).
\end{align}
In the Majorana representation $\overline{\chi} \gamma^\mu \chi= 0, $ hence terms 
containing such contractions do not contribute. However, even for Dirac fermions  
for which $\bar\chi\gamma^\mu \chi \neq 0$ terms with such contractions cancel 
in the final result, and the  expansion is, up to an overall factor of 2, given by the 
very same expression  \eqref{actionexp}.  Hence the anomaly for a Dirac fermion is twice the
one for a Majorana fermion. From the Lagrangian density above it is 
then straightforward to read off the Feynman rules with up to two external
graviton lines. The relevant expressions are given in figure \ref{feyn}.~\footnote{Since we 
are working with Lorentzian signature it should be understood that we are using the 
usual $i\varepsilon$ prescription for the propagator, although we do not write this out explicitly.} 

\vspace*{4mm}
\begin{figure} 
 \begin{tikzpicture}
\draw (3,0) node{$2i \frac{ p\!\!\!/}{p^2}$};
\draw[thick] [->](-1.5,0)--(0,0);
\draw[thick]  (0,0)--(1.5,0); 
\draw (0,0.1) node[above]{$p$};
\end{tikzpicture}

\vspace*{4mm}

\begin{tikzpicture}[
        thick,
        level/.style={level distance=1.5cm},
        level 2/.style={sibling distance=2.6cm},
        level 3/.style={sibling distance=2cm}
    ]
        \draw (6.3,0) node{$-\frac{1}{8} i \left[ (p-q)_{(\mu} \gamma_{\nu)} - \eta_{\mu \nu} (p\!\!\!/ - q\!\!\!/) \right]$};
    \coordinate
        child[grow=right, level distance=0pt] {
            child {
                child {
                           edge from parent [aelectron] node[above] {$ q$}
                }
                child {
                                edge from parent [aelectron] node[above] {$ p$}
                }
                 edge from parent [photon] node[left, below] {$ \mu \nu$}
            }
    };
\end{tikzpicture}

\vspace*{8mm}

\begin{tikzpicture}[thick]
\draw (5.5,0.75) node{$\frac{3}{64} i \left[ (p - q)_{(\mu} \eta_{\nu)(\rho} \gamma_{\sigma)}  +(p - q)_{(\rho} \eta_{\sigma)(\mu} \gamma_{\nu)} \right] $};
 \draw (5.4,0.00) node{$- \frac{1}{32} i \left[\eta_{ \mu \nu} (p - q)_{(\rho}  \gamma_{\sigma)}  + \eta_{\rho \sigma} (p - q)_{(\mu}  \gamma_{\nu)} \right]$};
  \draw (6.7,-0.75) node{$+ \frac{1}{32} i \left[\eta_{ \mu \nu} \eta_{\rho \sigma}  - 2 \eta_{\mu (\rho} \eta_{\sigma) \nu} \right] (p\!\!\!/ - q\!\!\!/)+ \frac{1}{64} i (k_{\a} - l_{\a}) \gamma^{\a (\mu}{}_{ (\rho} \delta^{\nu)}_{\sigma)}$};
\draw[photon](-1,1)--(0,0);
\draw[photon](-1,-1)--(0,0);
\draw[electron](1,1,0)--(0,0); 
\draw[electron](1,-1,0)--(0,0);
\draw (-1,1) node[left]{$\mu \nu$};
\draw (-1,-1) node[left]{$\rho \sigma$};
\draw (-0.45,0.5) node[above]{$k$};
\draw (-0.4,-0.5) node[below]{$l$};
\draw (0.5,0.5) node[above]{$p$};
\draw (0.6,-0.5) node[above]{$q$};
\end{tikzpicture}

\vspace*{8mm}

\begin{tikzpicture}[
        thick,
        level/.style={level distance=1.5cm},
        level 2/.style={sibling distance=2.6cm},
        level 3/.style={sibling distance=2cm}
    ]
        \draw (8.7,0.5) node{$-\frac{1}{8} i \left[ (p-q)_{(\mu} \eta_{\nu)(\rho} \gamma_{\sigma))} + \eta_{\mu \nu} (p-q)_{(\rho} \gamma_{\sigma))} - 2 \, \eta_{\mu (\rho} \eta_{\sigma) \nu} (p\!\!\!/ - q\!\!\!/) \right]$};
        \draw (6.2,-0.5) node{$-\frac{1}{16} i \, k^{\a} \, \left[ \gamma_{\a \mu (\rho} \eta_{\sigma)\nu } + \gamma_{\a \nu (\rho} \eta_{\sigma)\mu } \right]$};
        \draw (1.5,0.4) node{$\rho \sigma$};
        \draw (1.5,0) circle [radius=0.17];
        \draw (1.35, - 0.14) -- (1.64, 0.14);
        \draw (1.34,  0.13) -- (1.63, - 0.13);
    \coordinate
        child[grow=right, level distance=0pt] {
            child {
                child {
                           edge from parent [aelectron] node[above] {$ q$}
                }
                child {
                                edge from parent [aelectron] node[above] {$ p$}
                }
                 edge from parent [photon] node[left, below] {$ \mu \nu$}
            }
    };
\end{tikzpicture}
\caption{\it Feynman rules for graviton-fermion interactions; the crossed vertex
              comes from the expansion of $T_{\mu\nu}$ to $\cO(h)$.}
\label{feyn}
\end{figure}

We are interested in the expectation value of the stress tensor at first and second order in the metric perturbation,
\begin{align}
\langle T_{\mu \nu}(x) \rangle &= \Big\langle T_{\mu \nu}(x) e^{i(S^{(1)} + S^{(2)} + \cdots)}\Big\rangle_{0}, \notag \\[5pt]
&=\Big\langle \Big( T^{(0)}_{\mu \nu}(x) +  T^{(1)}_{\mu \nu}(x) +  \dots \Big) \Big( 1 + i  S^{(1)} + \left( i S^{(2)} - \frac12 S^{(1)} S^{(1)} \right) + \dots \Big)\Big\rangle_{0} , \notag \\[3pt] 
&= i \Big\langle  T^{(0)}_{\mu \nu}(x) S^{(1)} \Big\rangle_{0} +  i \Big\langle T^{(1)}_{\mu \nu}(x)  S^{(1)} \Big\rangle_{0} + i \Big\langle  T^{(0)}_{\mu \nu}(x)  S^{(2)} \Big\rangle_{0} \notag \\[3pt]
& \hspace{70mm} - \frac12 \Big\langle T^{(0)}_{\mu \nu}(x) S^{(1)} S^{(1)} \Big\rangle_{0} + \dots . \label{Texp}
\end{align}
where $\langle\cdots\rangle_0$ denotes the free expectation value (to be evaluated in the
spin-$\frac12$ Fock space). Note that at zeroth order, $\langle  T^{(0)}_{\mu \nu}(x) \rangle_{0}$, we only have tadpole diagrams, which vanish in dimensional regularisation. Furthermore, there 
is no $\langle  T^{(1)}_{\mu \nu}(x) \rangle_{0}$ contribution at first order in $h$, since these also  
contribute only tadpole diagrams.

\subsection{Expectation value of the stress tensor at $\mathcal{O}(h)$} \label{sub:1order}

In this section we briefly summarise the old $\cO(h)$ result of \cite{CapDuff}.
At first order, from equation \eqref{Texp} the expectation value of the stress tensor is 
\begin{align}
\big\langle T_{\mu \nu}(x) \big\rangle |_{\mathcal{O}(h)} =i \Big\langle  T^{(0)}_{\mu \nu}(x) S^{(1)} \Big\rangle_{0} = \int d^d y  \int\frac{d^dp}{(2 \pi)^d}  e^{-i p\cdot (x - y)}   T_{\mu\nu \rho \sigma } (p)h^{\rho \sigma}(y), \label{1hexpT}
\end{align}
which defines the two-point function $T_{\mu\nu \rho \sigma } (p)$ in momentum space.
Using the Feynman rules we have
\begin{align}
T_{\mu \nu \rho \sigma}(p) = \frac{i}{8} \int \frac{d^d k}{(2 \pi)^d} \textrm{tr} \left( \frac{k\!\!\!/}{k^2} (2 k - p)_{(\mu} \gamma_{\nu)} \frac{k\!\!\!/ - p\!\!\!/}{(k-p)^2} (2 k - p)_{(\rho} \gamma_{\sigma)} \right), \label{2pt}
\end{align}
where we have neglected all terms proportional to $\eta_{\mu \nu}$ and $\eta_{\rho \sigma}$, since, using the identity
\begin{equation}
  \frac{k\!\!\!/(2 k\!\!\!/ - p\!\!\!/)(k\!\!\!/ - p\!\!\!/ )}{k^2(k-p)^2} = \frac{(k\!\!\!/ - p\!\!\!/ )}{(k-p)^2} + \frac{k\!\!\!/}{k^2},
  \label{ident}
\end{equation}
these terms reduce to tadpole integrals which vanish. Note also the simple identities 
\beq
T_{\mu\nu\rho\sigma}(p) = T_{\mu\nu\rho\sigma}(-p) \;\;, \quad
T_{\mu\nu\rho\sigma}(p) = T_{\rho\sigma\mu\nu}(p).
\eeq
As shown in appendix \ref{app:subanti2pt}, equation \eqref{anti2pt},  the explicit symmetrisation 
of the $\mu \nu$ indices in the integral (\ref{2pt}) is not required, as the antisymmetric part 
vanishes, a fact that we will exploit to simplify some of the subsequent calculations.

The conservation of the stress tensor
\begin{equation}\label{pT1}
 \nabla^{\mu} \big\langle T_{\mu \nu} \big\rangle =0
\end{equation}
and the tracelessness
\begin{equation}\label{gT1}
\big\langle g^{\mu \nu}  T_{\mu \nu} \big\rangle =0
\end{equation}
 at order $h$, translate to the following Ward identities
\begin{align}
 p^{\mu} T_{\mu \nu \rho \sigma} &= 0, \label{1cons} \\[2pt]
  \eta^{(d) \, \mu \nu}T_{\mu \nu \rho \sigma} &= 0,\label{1trace}
\end{align}
where it is important that the trace {\em is taken in $d$ dimensions}
(indicated in the notation by putting the trace {\em inside} the brackets in (\ref{gT1}) and superscript $(d)$ on the $\eta$).
In order to verify the conservation Ward identity, we note that 
\begin{equation}\label{identity1}
 p^{\mu} (2 k - p)_{\mu} = k^2 - (k-p)^2.
\end{equation}
Hence $p^{\mu} T_{\mu \nu \rho \sigma}$ reduces to a tadpole integral which vanishes. 
Similarly, the $d$ dimensional trace reduces to a tadpole integral upon using identity \eqref{ident}.
This is in accord with the fact that the Dirac Lagrangian density is classically Weyl 
invariant in all dimensions with a $d$-dependent scaling of the fermions.

Evaluating the 2-point function integral, \eqref{2pt}, using the integral identities \eqref{2int1}--\eqref{2int4}, we obtain
\begin{gather} \label{tif}
- i \frac{1}{(2\sqrt{\pi})^d} \cdot
 \frac{2^{d/2} I(p)}{16 (d^2 - 1)} \bigg[ (d-2) p_{\mu} p_{\nu} p_{\rho} p_{\sigma} - 2 (d-1) p_{(\mu} \eta_{\nu)(\rho} p_{\sigma)} + p^2 \left( \eta_{\mu \nu} p_{\rho} p_{\sigma} +  \eta_{\rho \sigma} p_{\mu} p_{\nu}  \right) \notag\\[3pt]
  \hspace{50mm} + (p^2)^2 \left( (d-1) \eta_{\mu ( \rho} \eta_{\sigma) \nu} - \eta_{\mu \nu} \eta_{\rho \sigma} \right) \bigg]  .
 \end{gather}
where the extra factor of $1/(2\sqrt{\pi})^d$ in front is due to our normalisation of the integral $I(p)$ in (\ref{I(p)}). 
It is now straightforward to verify that the contraction of the above expression with $p^{\mu}$ vanishes, 
confirming that the Ward identity for general covariance is satisfied. Furthermore, we 
can verify again that the contraction of the $\mu \nu$ indices in $d$ dimensions is also zero.

However, contracting the $\mu \nu$ indices in four dimensions we obtain
 \begin{equation} \label{2pttrace}
  \eta^{(4) \, \rho\sigma} T_{\rho \sigma \mu \nu}
   =  \frac{p^2}{30 (4 \pi)^2} \left( p_{\mu} p_{\nu} - \eta_{\mu \nu} p^2 \right),
 \end{equation}
 from which we find the $\Box R$ anomaly at $\cO(h)$, to wit,
\beq
g^{\mu\nu} \langle T_{\mu\nu} \rangle \big|_{\cO(h)} =
\frac1{30 (4\pi)^2} \Box R \big|_{\cO(h)} \label{anom1oder}
\eeq
where we now put the $g^{\mu\nu}$ {\em outside} the bracket to indicate that the
trace is to be taken in four dimensions, after regularisation and renormalisation.

We stress that this $\cO(h)$ calculation can {\em not} give the $a$ and $c$ coefficients
as these require at least $\cO(h^2)$. However, with an extra assumption on the counterterm 
it is possible to derive the $c$ coefficient at least by indirect arguments \cite{Duff1}. This can be 
seen as follows: introducing the counterterm $\epsilon^{-1} \Delta W$, where 
$\Delta W \equiv \int d^dx \sqrt{-g} C^2$ and $C$ is the Weyl tensor,
and functionally differentiating, we get
\beq
2e^{-1} g_{\mu\nu} \frac{\delta}{\delta g_{\mu\nu}} \Delta W = (d-4)\left(C^2 + \frac23 \Box R \right)
\eeq
which shows that 
\beq\label{bc}
b = \frac23 c \; ,
\eeq
a relation which we shall later verify explicitly at $\cO(h^2)$. By contrast, there is no such 
indirect and labor saving argument for the coefficient $a$.

\subsection{Expectation value of the stress tensor at $\mathcal{O}(h^2)$} \label{sub:2order}

\vspace*{4mm}

\begin{figure}
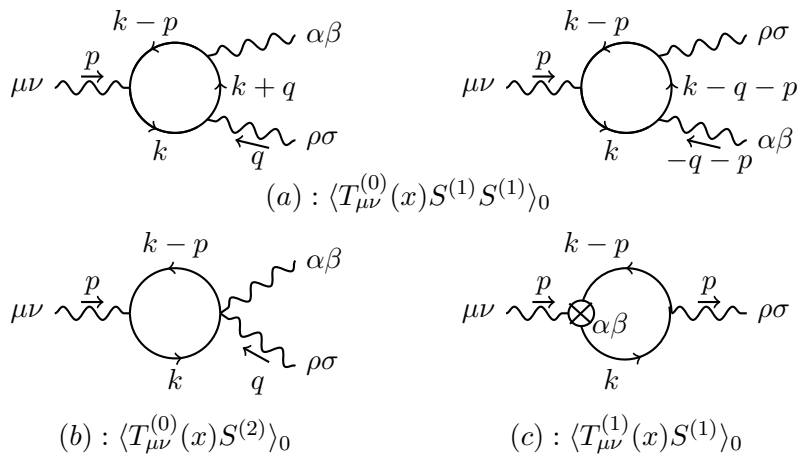

\tikz[scale=1,thick,doppelstriche/.style={postaction={decorate},
decoration={markings,
mark=at position .4 with {\draw (0,-3pt)--(0,3pt);},
mark=at position .6 with {\draw (0,-3pt)--(0,3pt);}
}}
,mdoppelstriche/.style={postaction={decorate},
decoration={markings,
mark=at position .35 with {\draw (0,-3pt)--(0,3pt);},
mark=at position .65 with {\draw (0,-3pt)--(0,3pt);}
}}]
{\pgfsetbaseline{-3pt};

\draw (-1,0) node[left]{};

\draw (1,0) node[left]{$\mu \nu$};
\draw (1.5,0.1) node[above]{$p$};
\draw [->] (1.35,0.15) -- (1.65,0.15);
\draw[photon](1,0)--(2,0);
\draw [aelectron] (2,0) arc [radius=0.6, start angle=180, end angle= 60];
\draw [electron] (2,0) arc [radius=0.6, start angle=180, end angle= 300];
\draw [electron] (2,0) arc [radius=0.6, start angle=180, end angle= 520];
\draw (4.2,0.7) node[right]{$\a \b$};
\draw (4.2,-0.7) node[right]{$\rho \sigma$};
\draw (2.2, 0.55) node[above]{$k -p$};
\draw (2.4, - 0.55) node[below]{$k$};
\draw (3.2, 0) node[right]{$k+q$};
\draw[photon](4.2,0.7,0)--(3.02,0.42); 
\draw[photon](4.2,-0.7,0)--(3.02,-0.42);
\draw (3.7, -0.7) node[below]{$q$};
\draw [->] (3.85, -0.8) -- (3.4,-0.693);

\draw (7,0) node[left]{$\mu \nu$};
\draw (7.5,0.1) node[above]{$p$};
\draw [->] (7.35,0.15) -- (7.65,0.15);
\draw[photon](7,0)--(8,0);
\draw [aelectron] (8,0) arc [radius=0.6, start angle=180, end angle= 60];
\draw [electron] (8,0) arc [radius=0.6, start angle=180, end angle= 300];
\draw [electron] (8,0) arc [radius=0.6, start angle=180, end angle= 520];
\draw (10.2,0.7) node[right]{$\rho \sigma$};
\draw (10.2,-0.7) node[right]{$\a \b$};
\draw (8.2, 0.55) node[above]{$k -p$};
\draw (8.4, - 0.55) node[below]{$k$};
\draw (9.2, 0) node[right]{$k-q-p$};
\draw[photon](10.2,0.7,0)--(9.02,0.42); 
\draw[photon](10.2,-0.7,0)--(9.02,-0.42);
\draw (9.7, -0.7) node[below]{$-q-p$};
\draw [->] (9.85, -0.8) -- (9.4,-0.693);

\draw (5.7,-1.8) node[above]{$(a): \langle T^{(0)}_{\mu \nu}(x) S^{(1)} S^{(1)} \rangle_{0} $};

\vspace{8mm}

\draw (1,-3) node[left]{$\mu \nu$};
\draw (1.5,-2.9) node[above]{$p$};
\draw [->] (1.35,-2.85) -- (1.65,-2.85);
\draw[photon](1,-3)--(2,-3);
\draw [electron] (3.2,-3) arc [radius=0.6, start angle=0, end angle= 180];
\draw [electron] (2,-3) arc [radius=0.6, start angle=180, end angle= 360];
\draw (4.2,-2.3) node[right]{$\a \b$};
\draw (4.2,-3.7) node[right]{$\rho \sigma$};
\draw (2.6, -2.4) node[above]{$k -p$};
\draw (2.6, - 3.6) node[below]{$k$};
\draw[photon](4.2,-2.3)--(3.2,-3); 
\draw[photon](4.2,-3.7)--(3.2,-3);
\draw (3.7, -3.7) node[below]{$q$};
\draw [->] (3.85, -3.65) -- (3.5,-3.455);

\draw (2.6,-5) node[above]{$(b): \langle T^{(0)}_{\mu \nu}(x) S^{(2)} \rangle_{0} $};

\draw (7,-3) node[left]{$\mu \nu$};
\draw (7.5,-2.9) node[above]{$p$};
\draw [->] (7.35,-2.85) -- (7.65,-2.85);
\draw[photon](7,-3)--(7.83,-3);
\draw [electron] (9.2,-3) arc [radius=0.6, start angle=0, end angle= 165];
\draw [electron] (8,-3.17) arc [radius=0.6, start angle=195, end angle= 370];
\draw (10.2,-3) node[right]{$\rho \sigma$};
\draw (8.2, -2.4) node[above]{$k -p$};
\draw (8.4, - 3.6) node[below]{$k$};
\draw (9.7, -2.9) node[above]{$p$};
\draw (8,-3.17) node[right]{$\a \b$};
\draw [->] (9.55,-2.85) -- (9.85,-2.85);
\draw[photon](9.2,-3)--(10.2,-3); 
        \draw (8 ,-3) circle [radius=0.17];
        \draw (7.85 , - 3.14) -- (8.14 , -2.86);
        \draw (7.84 , -2.87) -- (8.13 , - 3.13);
        \draw (8.6,-5) node[above]{$(c):\langle T^{(1)}_{\mu \nu}(x) S^{(1)}\rangle_{0} $};
}
\caption{\it Feynman diagrams for 3-point function of stress tensor insertions. The diagrams at the top, labelled (a), contribute to $T^{(1)}_{\mu \nu \rho \sigma \a \b}$; diagram (b) and (c) contribute to $T^{(2)}_{\mu \nu \rho \sigma \a \b}$ and  $T^{(3)}_{\mu \nu \rho \sigma \a \b}$, respectively. }
\label{3ptfey}
\end{figure}

The evaluation of the expectation value of the stress tensor
to second order in the metric fluctuations is far more involved than at first order
because there are many more contributions. In particular we must now consider
the 3-point functions which are given by Feynman diagrams labelled $(a)$ and $(b)$
in figure \ref{3ptfey}. These diagrams, as well as a new diagram labelled $(c)$ in figure \ref{3ptfey}, contribute to the the expectation value of the 
stress tensor at second order in the metric perturbation,
\begin{align}
\langle T_{\mu \nu}(x) \rangle |_{\mathcal{O}(h^2)} &= \int d^d y  \,  d^dz \int\frac{d^dp}{(2 \pi)^d}  \frac{d^dq}{(2 \pi)^d} e^{-i p\cdot (x - y) -i q\cdot (z - y)} h^{\rho \sigma}(y) \Bigg( T^{(1)}_{\mu\nu \rho \sigma \a \b} (p, q) \,  h^{\a \b}(z) \notag \\[5pt]
  & \hspace{40mm}+ T^{(2)}_{\mu\nu \rho \sigma \a \b} (p,q) \,  h^{\a \b}(z)  + T^{(3)}_{\mu\nu \rho \sigma \a \b} (p) \,  h^{\a \b}(x) \Bigg), \label{3ptfeyt}
\end{align}
Here $T^{1}_{\mu\nu \rho \sigma \a \b}$, $T^{(2)}_{\mu\nu \rho \sigma \a \b}$ and $T^{(3)}_{\mu\nu \rho \sigma \a \b}$ are the 3-point diagrams corresponding to $\langle T^{(0)}_{\mu \nu}(x) S^{(1)} S^{(1)} \rangle_{0} $, $\langle T^{(0)}_{\mu \nu}(x) S^{(2)} \rangle_{0} $ and $\langle T^{(1)}_{\mu \nu}(x) S^{(1)}\rangle_{0}$ in equation \eqref{Texp}, respectively.
Using the Feynman rules given in figure \ref{feyn}, the Feynman diagrams labelled (a) in figure \ref{3ptfey} give 
\begin{align}
 T^{(1)}_{\mu \nu \rho\sigma\a\b} (p,q) &= \frac{i}{64} \int \frac{d^dk}{(2 \pi)^d} \, \textrm{tr}\biggl\{ \frac{k\!\!\!/}{k^2} \left[ (2 k - p)_{(\mu} \gamma_{\nu)} - \eta_{\mu \nu} (2 k\!\!\!/ - p\!\!\!/ )\right] \frac{k\!\!\!/ - p\!\!\!/ }{(k-p)^2} \notag  \\[8pt]
 & \qquad \times \biggl( \left[ (2 k- p + q )_{(\rho} \gamma_{\sigma)} - \eta_{\rho \sigma} (2 k\!\!\!/ - p\!\!\!/ + q\!\!\!/ ) \right]
  \frac{k\!\!\!/ + q\!\!\!/ }{(k+q)^2}   \notag  \\[5pt]
 & \hspace{40mm}  \times \left[  (2k + q )_{(\a} \gamma_{\b)}  - \eta_{\a \b} (2 k\!\!\!/ + q\!\!\!/ ) \right] \notag \\[8pt]
 &  \qquad \quad + \left[  (2k - 2 p - q )_{(\a} \gamma_{\b)}  - \eta_{\a \b} (2 k\!\!\!/ - 2 p\!\!\!/ - q\!\!\!/ ) \right] 
  \frac{k\!\!\!/  - p\!\!\!/  - q\!\!\!/ }{(k- p- q)^2}  \notag  \\[5pt]
 & \hspace{40mm}  \times \left[ (2 k- p - q )_{(\rho} \gamma_{\sigma)} - \eta_{\rho \sigma} (2 k\!\!\!/ - p\!\!\!/ - q\!\!\!/ ) \right]\biggr)\biggr\}.  \label{Tp31}
\end{align}
Letting $k \rightarrow -k + p$ in the second term and using  the gamma matrix identity \eqref{gamma6id},
we can show that the second term is identical to the first term, \emph{viz}.\ the contribution from the two diagrams labelled (a) is identical. Furthermore the terms involving Kronecker $\delta$-symbols, can be written as two-point function integrals, defined in equation \eqref{2pt}, using identities analogous to \eqref{ident}. Terms with more than one Kronecker $\delta$-symbols reduce to tadpole integrals, which vanish, or integrals of the form 
\begin{equation}
\int \frac{d^d k}{(2 \pi)^d} \textrm{tr} \left( \frac{k\!\!\!/}{k^2} (2 k\!\!\!/ - p\!\!\!/ + q\!\!\!/ )  \frac{k\!\!\!/ - p\!\!\!/}{(k-p)^2} (2 k - p)_{(\rho} \gamma_{\sigma)} \right),
\end{equation}
which using identity \eqref{ident} reduces to 
\begin{equation}
\int \frac{d^d k}{(2 \pi)^d} \textrm{tr} \left( \frac{k\!\!\!/}{k^2} q\!\!\!/  \frac{k\!\!\!/ - p\!\!\!/}{(k-p)^2} (2 k - p)_{(\rho} \gamma_{\sigma)} \right) =0
\end{equation}
by identity \eqref{niceid}. Hence, we can rewrite $T^{(1)}$ as
\begin{align}
 T^{(1)}_{\mu \nu \rho\sigma\a\b} (p,q) &= 
 \frac{1}{32} T_{\mu \nu \rho \sigma \a \b}(p,q)
 -\frac14 \eta_{\mu \nu} \left[T_{\rho \sigma \a \b} (p+q) + T_{\rho \sigma \a \b} (q) \right] \notag \\[5pt]
 & \quad  -\frac14 \eta_{\rho \sigma} \left[T_{\mu \nu \a \b} (p) + T_{\mu \nu \a \b} (q) \right] -\frac14 \eta_{\a \b} \left[T_{\mu \nu\rho \sigma} (p+q) + T_{\mu \nu\rho \sigma} (p) \right], \label{T31}
\end{align}
where we define
\begin{equation}
T_{\mu \nu \rho \sigma \a \b} \equiv i \int \frac{d^dk}{(2 \pi)^d} \, \textrm{tr}\biggl\{ \frac{k\!\!\!/}{k^2} (2 k - p)_{(\mu} \gamma_{\nu)}  \frac{k\!\!\!/ - p\!\!\!/ }{(k-p)^2} (2 k- p + q )_{(\rho} \gamma_{\sigma)}
 \frac{k\!\!\!/ + q\!\!\!/ }{(k+q)^2}  (2k + q )_{(\a} \gamma_{\b)} \biggr\}.
\end{equation}
that is, the original expression but {\em without} the trace terms.

Moreover, the Feynman diagrams labelled (b) and (c), respectively, give
\begin{align}
 T^{(2)}_{\mu \nu \rho\sigma\a\b} (p,q) &= - \frac{i}{64} \int \frac{d^dk}{(2 \pi)^d} \, \textrm{tr}\biggl\{ \frac{k\!\!\!/}{k^2} (2 k - p)_{(\mu} \gamma_{\nu)} \frac{k\!\!\!/ - p\!\!\!/ }{(k-p)^2} \notag  \\[8pt]
 & \quad \times \biggl( 3 \left( \eta_{(\rho|(\a} (2 k- p )_{\b)} \gamma_{|\sigma)} + \eta_{(\a|(\rho} (2 k- p )_{\sigma)} \gamma_{|\b)} \right)\notag  \\[5pt]
 &  \qquad \quad -2 \left( \eta_{\rho \sigma} (2 k- p )_{(\a} \gamma_{\b)} + \eta_{\a \b} (2 k- p )_{(\rho} \gamma_{\sigma)}\right)\notag  \\[5pt]
 &  \qquad \quad + \frac12 (2 q + p)^{\tau} \left(  \gamma_{\tau \a (\rho} \eta_{\sigma) \b} + \gamma_{\tau \b (\rho} \eta_{\sigma) \a} \right)   \biggr)\biggr\},  \label{Tp32}
\end{align}
\begin{align}
 T^{(3)}_{\mu \nu \rho\sigma\a\b} (p) &= \frac{i}{16} \int \frac{d^dk}{(2 \pi)^d} \, \textrm{tr}\biggl\{  \frac{k\!\!\!/ - p\!\!\!/ }{(k-p)^2} (2 k- p  )_{(\rho} \gamma_{\sigma)}\frac{k\!\!\!/}{k^2} \notag  \\[8pt]
 & \hspace{5mm} \times \left[ (2 k - p)_{(\mu} \eta_{\nu)(\a} \gamma_{\b)} +  \eta_{\mu \nu} (2 k - p)_{(\a} \gamma_{\b)} + \frac12 p^{\tau} \left( \gamma_{\tau \a (\mu } \eta_{\nu) \b} + \gamma_{\tau \b (\mu } \eta_{\nu) \a} \right) \right] \biggr\},  \label{Tp33}
 \end{align}
 where, as in the two-point function evaluation, we have used the fact that some terms lead to tadpole integrals which vanish. It is also straightforward to show that the terms proportional to $\gamma^{ \tau \a \b}$ in both $T^{(2)}$ and $T^{(3)}$ vanish. Therefore, we can express both contributions solely in terms of the two-point function integral, \eqref{2pt},
 \begin{align}
 T^{(2)}_{\mu \nu \rho\sigma\a\b} (p) &= - \frac34 \eta_{(\a|(\rho} T_{\sigma)|\b) \mu \nu} (p) + \frac14 \eta_{\rho \sigma} T_{\mu \nu \a \b}(p) + \frac14 \eta_{\a \b} T_{\mu \nu\rho \sigma} (p),  \label{T32}\\[5pt]
 T^{(3)}_{\mu \nu \rho\sigma\a\b} (p) &= \frac12 \eta_{(\a|(\mu} T_{\nu)|\b) \rho \sigma} (p) + \frac12 \eta_{\mu \nu} T_{\rho \sigma \a \b}(p).  \label{T33}
\end{align}

\subsection{Conservation} \label{sub:con}

The conservation of the expectation value of the stress tensor can be expressed 
as follows:
\begin{equation}
 \nabla^{\mu} \langle T_{\mu \nu} \rangle = g^{\mu \rho} \Big( \partial_{\rho}  \langle T_{\mu \nu} \rangle - \Gamma^{\sigma}_{\rho \mu} \langle T_{\sigma \nu} \rangle - \Gamma^{\sigma}_{\rho \nu} \langle T_{\mu\sigma } \rangle \Big)=0 . \label{covcons}
\end{equation}
Using the expansion of the metric and Christoffel symbol components in appendix \ref{app:exp}, at second order in the metric perturbation the above identity reduces to 
\begin{equation}
\partial^{\mu}  \langle T_{\mu \nu} \rangle|_{\mathcal{O} (h^2)} - h^{\mu \rho} \partial_{\rho} \langle T_{\mu \nu} \rangle|_{\mathcal{O}(h)}  - \frac12 (2 \partial_{\mu} h^{\mu \rho} - \partial^{\rho} h) \langle T_{\rho \nu} \rangle|_{\mathcal{O}(h)}  - \frac12 \partial_{\nu} h^{\mu \rho} \langle T_{\mu\rho } \rangle|_{\mathcal{O}(h)} =0, \label{cons2order}
\end{equation}
where $ \langle T_{\mu \nu} \rangle|_{\mathcal{O}(h)}$ and 
$ \langle T_{\mu \nu} \rangle|_{\mathcal{O}(h^2)}$ are defined in equations \eqref{1hexpT} and \eqref{3ptfeyt}, respectively. 
Equation \eqref{covcons} is fully covariant. However, in equation \eqref{cons2order}, and for the rest of the section, the indices are raised and lowered with the flat metric. 
 
We first consider 
\begin{align}
 \partial^{\mu}  \langle T_{\mu \nu} \rangle(x)\big|_{\mathcal{O}(h^2)} &=  \, \int d^d y  \,  d^dz \int\frac{d^dp}{(2 \pi)^d}  \frac{d^dq}{(2 \pi)^d} e^{-i p\cdot (x - y) -i q\cdot (z - y)} h^{\rho \sigma}(y)  \notag \\[5pt]
  & \hspace{5mm} \times \Biggr\{ (- ip^{\mu}) \, \Bigg( T^{(1)}_{\mu\nu \rho \sigma \a \b} (p, q) \,  h^{\a \b}(z)+ T^{(2)}_{\mu\nu \rho \sigma \a \b} (p,q) \,  h^{\a \b}(z) \notag \\[5pt]
 & \hspace{35mm} + T^{(3)}_{\mu\nu \rho \sigma \a \b} (p) \,  h^{\a \b}(x) \Bigg) +  T^{(3)}_{\mu\nu \rho \sigma \a \b} (p) \, \partial^{\mu} h^{\a \b}(x) \Biggr\}. \label{pTp}
\end{align}
Using equation \eqref{T31} and the conservation Ward identity at first order in $h$, \eqref{1cons},
\begin{align}
 p^{\mu} \, T^{(1)}_{\mu\nu \rho \sigma \a \b} (p, q) &= \frac{1}{32} \, p^{\mu} \,  T_{\mu \nu \rho \sigma \a \b} -\frac14 p_{\nu} \left[T_{\rho \sigma \a \b} (p+q) + T_{\rho \sigma \a \b} (q) \right] \notag \\[5pt]
 & \quad \, -\frac14  p^{\mu} \left[ \eta_{\rho \sigma} T_{\mu \nu \a \b} (q) +  \eta_{\a \b} T_{\mu \nu\rho \sigma} (p+q) \right].  \label{T31pcons}
\end{align}
Furthermore, using equations \eqref{T32} and \eqref{T33},
 \begin{align}
  p^{\mu} \, T^{(2)}_{\mu \nu \rho\sigma\a\b} (p) &= 0,  \label{T32cons}\\[5pt]
  p^{\mu} \, T^{(3)}_{\mu \nu \rho\sigma\a\b} (p) &= \frac14 \, p_{(\a} \, T_{\b) \nu \rho \sigma} (p) + \frac12 \, p_{\nu}\,  T_{\rho \sigma \a \b}(p).  \label{T33cons}
\end{align}
We have expressed all the terms in terms of the two-point function integral except the first term on the r.h.s.\  of equation \eqref{T31pcons}, which we would like to also rewrite in terms of two-point function integral,
\begin{align}
& p^{\mu} \,T_{\mu \nu \rho \sigma \a \b} \notag \\[3pt]
=& \; \frac{i}{2} \int \frac{d^dk}{(2 \pi)^d} \, \textrm{tr}\biggl\{ \Big( k\!\!\!/ \gamma_{\nu} + (2 k- p)_{\nu} \Big) \frac{k\!\!\!/ - p\!\!\!/ }{(k-p)^2} (2 k- p + q )_{(\rho} \gamma_{\sigma)} \frac{k\!\!\!/ + q\!\!\!/ }{(k+q)^2}  (2k + q )_{(\a} \gamma_{\b)} \biggr\} \notag \\[3pt]
&  -  \frac{i}{2} \int \frac{d^dk}{(2 \pi)^d} \, \textrm{tr}\biggl\{ \frac{k\!\!\!/ }{k^2} \Big(  \gamma_{\nu} (k\!\!\!/ - p\!\!\!/ ) + (2 k- p)_{\nu} \Big)  (2 k- p + q )_{(\rho} \gamma_{\sigma)} \frac{k\!\!\!/ + q\!\!\!/ }{(k+q)^2}  (2k + q )_{(\a} \gamma_{\b)} \biggr\},
\end{align}
where we have used (\ref{identity1}) and $p\!\!\!/ = k\!\!\!/ - (k\!\!\!/ - p\!\!\!/ )$ to cancel a 
propagator factor. We redefine $k \rightarrow -k + p$ in the first term and $k \rightarrow -k$ 
in the second term, whereupon we obtain 
\begin{equation}
 p^{\mu} \,T_{\mu \nu \rho \sigma \a \b} = \frac12 \Big( \tilde{T}_{\nu \rho \sigma \a \b}(p,q) - \tilde{T}_{\nu \a \b\rho \sigma }(-p, p+q) \Big),
\end{equation}
where 
  \begin{align}
 \tilde{T}_{\nu \rho \sigma \a \b}(p,q) & \equiv i \int  \frac{d^dk}{(2 \pi)^d} \, \textrm{tr}\biggl\{ \frac{k\!\!\!/ }{k^2} \Big(  \gamma_{\nu} (k\!\!\!/ + p\!\!\!/ ) + (2 k + p)_{\nu} \Big)  (2 k + p - q )_{(\rho} \gamma_{\sigma)}  \frac{k\!\!\!/ - q\!\!\!/ }{(k-q)^2}  (2k - q )_{(\a} \gamma_{\b)} \biggr\} \label{Ttildedef}
\end{align}
and we have used the identity \eqref{gamma6id}. In appendix \ref{app:Ttilde}, we simplify
this integral and derive equation \eqref{int1ans}. Using this result, we arrive at
\begin{align}
 p^{\mu} \, T_{\mu \nu \rho \sigma \a \b}
&= \, 4 \Big[ 3 \, p_{(\rho}\,  T_{\sigma) \nu \a \b}(q) + 2\, (p+ q)_{\nu} \, T_{ \rho \sigma \a \b}(q) - p^{\tau} \, \eta_{\nu (\rho} \,  T_{\sigma) \tau \a \b}(q)  \notag \\[5pt]
& \hspace{10mm}  +3 \, p_{(\a}\,  T_{\b) \nu \rho \sigma}(p+q) - 2\, q_{\nu} \, T_{ \rho \sigma \a \b}(p+q) - p^{\tau} \, \eta_{\nu (\a} \,  T_{\b) \tau \rho \sigma}(p+q) \Big].
\end{align}
Hence, from equation \eqref{T31pcons}, 
\begin{align}
 p^{\mu} \, T^{(1)}_{\mu\nu \rho \sigma \a \b} (p, q) &= \,  
 \frac18 \Big[  3 \, p_{(\a}\,  T_{\b) \nu \rho \sigma}(p+q) - 2\,  (p+q)_{\nu} \, T_{ \rho \sigma \a \b}(p+q) - 2\,  p^{\mu} \eta_{\a \b} T_{\mu \nu\rho \sigma} (p+q)  \notag \\[3pt]
& \qquad \quad  - p^{\tau} \, \eta_{\nu (\a} \,  T_{\b) \tau \rho \sigma}(p+q) + 3 \, p_{(\rho}\,  T_{\sigma) \nu \a \b}(q) + 2\,  q_{\nu} \, T_{ \rho \sigma \a \b}(q) \notag \\[5pt]
 & \qquad \quad -2 \, p^{\mu}  \eta_{\rho \sigma} T_{\mu \nu \a \b} (q)  - p^{\tau} \, \eta_{\nu (\rho} \,  T_{\sigma) \tau \a \b}(q) \Big].  \label{T31cons}
\end{align}
Integrating the above equation over $p$ and letting $p \rightarrow p-q$,
\begin{align}
  \int\frac{d^dp}{(2 \pi)^d} e^{-i p\cdot (x - y)} p^{\mu} \, T^{(1)}_{\mu\nu \rho \sigma \a \b} (p, q) &=  \frac18 \int\frac{d^dp}{(2 \pi)^d} e^{-i (p - q)\cdot (x - y)} \,  \left[  3 \, (p-q)_{(\a}\,  T_{\b) \nu \rho \sigma}(p)  \right.\notag \\[5 pt]
& \quad  - 2\,  p_{\nu} \, T_{ \rho \sigma \a \b}(p) + 2\,  q^{\mu} \eta_{\a \b} T_{\mu \nu\rho \sigma} (p)  + q^{\tau} \, \eta_{\nu (\a} \,  T_{\b) \tau \rho \sigma}(p) \notag \\[5pt]
& \quad + 3 \, (p-q)_{(\rho}\,  T_{\sigma) \nu \a \b}(q)  + 2\,  q_{\nu} \, T_{ \rho \sigma \a \b}(q)  \notag \\[5pt]
 & \quad-2 \, p^{\mu} \left. \eta_{\rho \sigma} T_{\mu \nu \a \b} (q)  - p^{\tau} \, \eta_{\nu (\rho} \,  T_{\sigma) \tau \a \b}(q) \right]. 
\end{align}
Therefore, using also equations \eqref{T32cons} and \eqref{T33cons} and \eqref{T33},
\begin{align}
 \partial^{\mu}  \langle T_{\mu \nu} \rangle(x)|_{\mathcal{O}(h^2)} &= \, \int d^d y  \,  d^dz \int\frac{d^dp}{(2 \pi)^d}  \frac{d^dq}{(2 \pi)^d} e^{-i p\cdot (x - y) -i q\cdot (z - x)} h^{\rho \sigma}(y)  \notag \\[5pt]
   & \qquad \times \Biggl\{ - \frac{i}{8}  h^{\a \b}(z)  \Big[  (5\, p - 3\, q)_{\a}\,  T_{\b \nu \rho \sigma}(p) + 2\,  p_{\nu} \, T_{ \rho \sigma \a \b}(p) \notag \\[5 pt]
& \qquad \quad  + 2\,  q^{\mu} \eta_{\a \b} T_{\mu \nu\rho \sigma} (p)  + q^{\tau} \, \eta_{\nu \a} \,  T_{\b \tau \rho \sigma}(p)+ 3 \, (p-q)_{\rho}\,  T_{\sigma \nu \a \b}(q)  \notag \\[5pt]
 & \qquad \quad + 2\,  q_{\nu} \, T_{ \rho \sigma \a \b}(q) -2 \, p^{\mu}  \eta_{\rho \sigma} T_{\mu \nu \a \b} (q)  - p^{\tau} \, \eta_{\nu \rho} \,  T_{\sigma \tau \a \b}(q) \Big] \notag \\
 & \qquad \qquad  + \frac14 \Big[ T_{\rho \sigma \nu \a} (p) \partial_{\b} h^{\a \b}(z) +  T_{\rho \sigma \mu \a} \eta_{\b \nu} (p) \partial_{\mu} h^{\a \b}(z) \notag \\
 & \hspace{60mm} + 2\,   T_{\rho \sigma \a \b} (p) \partial_{\nu} h^{\a \b}(z) \Big]\Biggr\}.
\end{align}
By reparametrising the integration variables, the terms in the integrand that are proportional to two-point function integrals with arguments $q$ can be replaced by terms proportional to  those with arguments $p$~\footnote{More precisely, the relabelling of the integration variables implies that the integrand must be invariant under $ p \leftrightarrow -q$ and $(\rho \sigma) \leftrightarrow (\a \b)$.}. Whereupon, 
\begin{align}
 \partial^{\mu}  \langle T_{\mu \nu} \rangle\big|_{\mathcal{O}(h^2)} &= - \frac14 \, \int d^d y  \,  d^dz \int\frac{d^dp}{(2 \pi)^d}  \frac{d^dq}{(2 \pi)^d} e^{-i p\cdot (x - y) -i q\cdot (z - x)} h^{\rho \sigma}(y)  \notag \\[5pt]
   & \qquad \times \Bigl( i\,  h^{\a \b}(z)  \Big[  (4\, p - 3\, q)_{\a}\,  T_{\b \nu \rho \sigma}(p)  + 2\,  q^{\mu} \eta_{\a \b} T_{\mu \nu\rho \sigma} (p) + \, q^{\tau} \, \eta_{\nu \a} \,  T_{\b \tau \rho \sigma}(p) \Big] \notag \\[8 pt]
 & \hspace{10mm}  -  T_{\rho \sigma \nu \a} (p) \partial_{\b} h^{\a \b}(z)- \eta^{\a \b} T_{\rho \sigma \mu \a} (p)  \partial^{\mu} h_{\b \nu}(z) - 2\,   T_{\rho \sigma \a \b} (p) \partial_{\nu} h^{\a \b}(z) \Big] \Big).
\end{align}
Integrating by parts over the $y$ and $z$ integrals,
\begin{align}
 \partial^{\mu}  \langle T_{\mu \nu} \rangle|_{\mathcal{O}(h^2)} &= \, \int d^d y  \,  d^dz \int\frac{d^dp}{(2 \pi)^d}  \frac{d^dq}{(2 \pi)^d} e^{-i p\cdot (x - y) -i q\cdot (z - x)} h^{\rho \sigma}(y) \Biggl\{  - i \, p_{\a}  T_{\b \nu \rho \sigma}(p)  h^{\a \b}(z)  \notag \\[5pt]
   & \qquad   +  \partial_{\a} h^{\a \b}(z)   T_{\b \nu \rho \sigma}(p) - \frac12 \,  \partial^{\mu} h(z) T_{\mu \nu\rho \sigma} (p) +  \frac 12 \partial_{\nu} h^{\a \b}(z) T_{\rho \sigma \a \b} (p)   \Biggr\}, \notag \\
   & \!\!\!\!\!\!\!\!\!\!\!\!\!\!\!\!\!\!\!\!\!\!\!\!
   = \, h^{\a \b} \partial_{\a} \langle T_{\b \nu} \rangle|_{\mathcal{O}(h)} + \partial_{\a} h^{\a \b} \langle T_{\b \nu} \rangle|_{\mathcal{O}(h)} - \frac12 \partial^{\mu} h \langle T_{\mu \nu} \rangle|_{\mathcal{O}(h)} + \frac12 \partial_{\nu} h^{\a \b}(z) \langle T_{\a \b} \rangle|_{\mathcal{O}(h)},
\end{align}
where in the last line we have integrated over $q$ and $z$, which sets $z=x$, and used definition \eqref{1hexpT}. We have, therefore, verified equation \eqref{cons2order} and hence
\begin{equation}
 \nabla^{\mu} \langle T_{\mu \nu} \rangle =0
\end{equation}
up to and including second order in the metric perturbation.

\section{Scalar 3-point loop integrals} \label{sec:3pointidentities}

We have seen in the preceding chapters that the evaluation of the expectation value of
the stress tensor to second order in the metric fluctuations requires the 
computation of certain 3-point Feynman  loop integrals (and correspondingly the
evaluation of $(n+1)$-point loop integrals if one expands to $n$-th order in the 
metric fluctuations). Such integrals have  been much investigated in the literature,
see e.g. \cite{Smirnov,Wein} for recent reviews and references. Nevertheless, and
also with regard to possible future applications, we here collect some formulae needed 
for our computation that to the best of our knowledge have not been given in fully 
explicit form in the literature, although the general procedure for their derivation is 
of course known, see in particular \cite{Davyd1,Davyd2,Boos}.

The relevant integrals  are of the form
\begin{equation} \label{tensorint}
  J_{\mu_1 \dots \mu_{M}} (d \,|\, p,q)= \int \frac{d^dk}{\pi^{d/2}} 
  \frac{k_{\mu_1}\cdots k_{\mu_{M}}}{k^2 (k - p)^2 (k +q)^2},
\end{equation}
or more generally
\begin{equation}\label{tensorint1}
 J_{\mu_1 \dots \mu_{M}}(d;m_1, m_2, m_3\,|\, p,q) \equiv \int \frac{d^dk}{\pi^{d/2}} \frac{k_{\mu_1}\cdots k_{\mu_{M}}}{k^{2m_1} (k - p)^{2m_2} (k +q)^{2m_3}}
\end{equation}
with (not necessarily integer) exponents $m_1,m_2,m_3$ \footnote{In the remainder we will 
usually not write out all arguments displayed on the l.h.s.\ of (\ref{tensorint1}).}. 
For the computation of the conformal anomaly we are 
in particular interested in the pole part of these integrals for $d \rightarrow 4$. Note that
we normalise the loop integrals (\ref{tensorint}) and (\ref{tensorint1}) with the factor 
$\pi^{-d/2}$, different from the normalisation adopted in the rest of this paper. This we
do only for convenience in order to simplify the subsequent calculations: because
\beq
\frac1{(2\pi)^d} = \frac1{(2\sqrt{\pi})^d}\cdot \frac1{\pi^{d/2}}
\eeq
we then only need to multiply the final results by $(2\sqrt{\pi})^{-d}$ to revert to the 
normalisation conventions used in the rest of this article. 

To evaluate the integrals we will follow a method developed by Davydychev \cite{Davyd1,Davyd2},
whereby the above integrals can be reduced to the basic scalar 3-point loop integral
 \begin{gather} \label{Jd111}
   J(d; 1,1,1\,|\,p,q ) 
   =  \int \frac{d^dk}{\pi^{d/2}} \, \frac{1}{k^{2} (k - p)^{2} (k + q)^{2} }
\end{gather}
which is again a special case of the more general integral
 \begin{gather} \label{Jdnu}
   J(d; m_1, m_2, m_3 \,|\, p,q) 
   =  \int \frac{d^dk}{\pi^{d/2}} \, \frac{1}{k^{2 m_1} (k - p)^{2 m_2} (k + q)^{2 m_3} }
\end{gather}
and so-called boundary integrals for which one of the exponents $m_i$ vanishes 
(see appendix \ref{app:int})
\begin{equation}\label{Ipd}
 I(d;m_1,m_2 \,|\, p) \equiv J(d;m_1,m_2,0\,|\, p) 
 =   \int \frac{d^dk}{\pi^{d/2}} \, \frac{1}{k^{2m_1} (p-k)^{2m_2}} 
\end{equation}
up to explicit factors which are rational functions of the external momenta.  
The final result will be completely explicit because for (\ref{Ipd})  we have 
the explicit formula
\begin{equation}\label{Ipd1}
\int \frac{d^dk}{\pi^{d/2}} \, \frac{1}{k^{2m_1} (p-k)^{2m_2}} \,=\, 
i (p^2)^{d/2-m} \frac{\Gamma(m - \frac{d}{2}) \Gamma(\frac{d}{2}-m_1)\Gamma(\frac{d}2 - m_2)}{\Gamma(d-m)\Gamma(m_1) \Gamma(m_2)}.
\end{equation}
where $m \equiv m_1 + m_2$ and the factor of $i$ comes from Wick rotating from Lorentzian space
to Euclidean signature. A further advantage of our choice is the simple normalisation 
\beq
I(p) \,\equiv \, I(d;1,1|p) \, = \, \frac{i}{\epsilon} \, + \, \cO(1)
\eeq
As we said, our derivation relies largely on the general formalism developed in \cite{Davyd1,Davyd2}  
but we  will spell out the formulae given there in more detail for the cases of interest. 
The final result will thus express (\ref{tensorint}) directly in terms of 
explicitly known functions, where all the UV divergences (needed for the determination
of the conformal anomaly) are contained in the boundary integrals. 
The extension of our results to higher $n$-point scalar loop integrals is
straightforward, though increasingly tedious for higher values of $n$.

In the remainder we will assume the external momenta $p$ and $q$ to assume generic values, 
for which $p^2 q^2  \neq \pq^2 $, so as to avoid IR or kinematical singularities -- the latter can then 
be easily and explicitly extracted from our final expressions. 
First we note that, in the Feynman parametrisation, the scalar integral (\ref{Jdnu}) is given by
\begin{equation}
J (d; m_1, m_2, m_3 ) = 
\frac{i \, \Gamma(m-\frac{d}{2})}{ \Gamma(m_1) \Gamma(m_2) \Gamma(m_3)} 
\int_{0}^{1}d \xi_1 d \xi_2 d \xi_3\,
  \frac{ \xi_1^{m_1 -1}\, \xi_2^{m_2 -1} \, \xi_3^{m_3 - 1 } \, 
\delta(1-\xi_1 - \xi_2 - \xi_3)}{\big[\xi_1\xi_2 p^2 + \xi_1\xi_3 q^2 + \xi_2 \xi_3(p+q)^2\big]^{m-d/2}}, 
\label{intfeypara}
\end{equation}
where $m \equiv m_1 + m_2 + m_3$.
Differentiating the l.h.s.\  of (\ref{Jdnu}) with respect to $p_{\mu}$ we find
\begin{equation} \label{dlhsintfeypara}
  2 \,m_2 \Big( J_{\mu}(d;m_1,m_2+1,m_3) \, -  \, p_{\mu} \, J(d;m_1,m_2+1,m_3) \Big),
\end{equation}
On the other hand, differentiating the r.h.s.\  of equation \eqref{intfeypara} gives 
\begin{equation} \label{drhsintfeypara}
 - 2 m_2 \Big( m_1 \, p_{\mu} J(d+2;m_1 + 1,m_2+1,m_3) 
   + m_3 \, (p_{\mu} +q_{\mu}) J(d+2;m_1,m_2+1,m_3 + 1) \Big).
\end{equation}
Equating expressions \eqref{dlhsintfeypara} and \eqref{drhsintfeypara}, we obtain an 
equation for $J_{\mu}(d;m_1, m_2, m_3)$ in terms of scalar integrals \cite{Davyd1}.
We can further simplify this expression by noting the identity
\begin{equation}
J(d;\{m_i\})= \sum_{j=1}^3  m_j \, J(d+2;\{m_i + \delta_{ij}\}),
\end{equation}
which can be proved directly from \eqref{intfeypara}.  Using the above identity and equating 
expressions \eqref{dlhsintfeypara} and \eqref{drhsintfeypara},  we obtain \cite{Davyd1}
\begin{equation}
J_{\mu}(d;m_1, m_2, m_3) = m_2 \, p_{\mu} \, J^{(3)}(d+2;m_1,m_2+1,m_3) - m_3 \, q_{\mu} \, J^{(3)}(d+2;m_1,m_2,m_3 + 1).
\end{equation}
This method can be inductively implemented, by further differentiating with respect to $p_{\mu}$, 
to find similar identities for $J_{\mu_1 \dots \mu_{M}}$ in terms of scalar integrals,
see \cite{Davyd2} for the general formulae. 
We list the relevant identities for $M$ up to $M=6$, found using the method outlined above
  \begin{align}
  J_{\mu}(d; 1,1,1) &= p_{\mu} \, J(d+2; 1, 2, 1) - q_{\mu} \, J(d+2; 1, 1, 2), \\[8pt]
  J_{\mu \nu}(d;1,1,1)  &= \frac{1}{2} \eta_{\mu \nu} \, J(d+2; 1, 1, 1 ) + 2 \Big[ p_{\mu} p_{\nu} \, J(d+4; 1, 3, 1 ) \notag \\[5pt]
  & \quad - p_{(\mu} q_{\nu)} \, J(d+4; 1, 2, 2 ) + q_{\mu} q_{\nu} \, J(d+4; 1, 1, 3 ) \Big], \label{3int2} \\[8pt]
  J_{\mu \nu \rho}(d;1,1,1)  &= \frac{3}{2}\eta_{(\mu \nu}  \Big[ p_{\rho)} \, J(d+4; 1, 2, 1 ) - q_{\rho)} \, J(d+4; 1, 1, 2 ) \Big] \notag \\[5pt] 
  &\quad + 6 \Big[ p_{\mu} p_{\nu} p_{\rho} \, J(d+6; 1, 4, 1 ) - p_{(\mu} p_{\nu} q_{\rho)} \, J(d+6; 1, 3, 2 ) \notag \\[5pt]
  & \qquad + p_{(\mu} q_{\nu} q_{\rho)} \, J(d+6; 1, 2, 3 ) - q_{\mu} q_{\nu} q_{\rho} \, J(d+6; 1, 1, 4 ) \Big], \notag \\[5pt]
     J_{\mu \nu \rho \sigma}(d;1,1,1)  &= \frac{3 }{4} \eta_{(\mu \nu} \eta_{\rho \sigma)}\, J(d+4; 1, 1, 1 ) + 6 \, \eta_{(\mu \nu} \Big[ p_{\rho} p_{\sigma)} \, J(d+6; 1, 3, 1 )\notag \\[5pt] 
     &\quad - p_{\rho} q_{\sigma)} \, J(d+6; 1, 2, 2 ) + q_{\rho} q_{\sigma)} \, J(d+6; 1, 1, 3 ) \Big] \notag \\[5pt]
     & \quad + 24 \Big[ p_{\mu} p_{\nu} p_{\rho} p_{\sigma} \, J(d+8; 1, 5, 1 ) - p_{(\mu} p_{\nu} p_{\rho} q_{\sigma)} \, J(d+8; 1, 4, 2 ) \notag \\[5pt] 
     & \qquad + p_{(\mu} p_{\nu} q_{\rho} q_{\sigma)} \, J(d+8; 1, 3, 3 ) - p_{(\mu} q_{\nu} q_{\rho} q_{\sigma)} \, J(d+8; 1, 2, 4 ) \notag \\[5pt] 
     & \qquad + q_{\mu} q_{\nu} q_{\rho} q_{\sigma} \, J(d+8; 1, 1, 5 ) \Big], \\[8pt]   
       J_{\mu \nu \rho \sigma \alpha}(d;1,1,1)  &= \frac{15}{4} \eta_{(\mu \nu} \eta_{\rho \sigma} \Big[ p_{\a)}  \, J(d+6; 1, 2, 1 )- q_{\a)}  \, J(d+6; 1, 1, 2 ) \Big] \notag \\[5pt] 
       &\quad +30\,  \eta_{(\mu \nu} \Big[ p_{\rho} p_{\sigma} p_{\a)} \, J(d+8; 1, 4, 1 )- p_{\rho} p_{\sigma} q_{\a)} \, J(d+8; 1, 3, 2 ) \notag \\[5pt] 
            &\qquad +  p_{\rho} q_{\sigma} q_{\a)} \, J(d+8; 1, 2, 3 ) - q_{\rho} q_{\sigma} q_{\a)} \, J(d+8; 1, 1, 4 ) \Big] \notag \\[5pt] 
       &\quad + 120 \Big[ p_{\mu}  p_{\nu} p_{\rho} p_{ \sigma} p_{\a} \, J(d+10; 1, 6, 1 ) -  p_{(\mu}  p_{\nu} p_{\rho} p_{ \sigma} q_{\a)} \, J(d+10; 1, 5, 2 ) \notag \\[5pt] 
        &\qquad +  p_{(\mu}  p_{\nu} p_{\rho} q_{ \sigma} q_{\a)} \, J(d+10; 1, 4, 3 ) -  p_{(\mu}  q_{\nu} q_{\rho} q_{ \sigma} q_{\a)} \, J(d+10; 1, 3, 4 ) \notag \\[5pt]
         &\qquad +  p_{(\mu}  q_{\nu} q_{\rho} q_{ \sigma} q_{\a)} \, J(d+10; 1, 2, 5 ) -  q_{\mu}  q_{\nu} q_{\rho} q_{ \sigma} q_{\a} \, J(d+10; 1, 1, 6 ) \Big],  
         \end{align}
         \begin{align}
  J_{\mu \nu \rho \sigma \a \b}(d;1,1,1)  &= \frac{15}{8} \eta_{(\mu \nu} \eta_{\rho \sigma} \eta_{\a \b)} \, J(d+6; 1, 1, 1 ) + \frac{45 }{2} \eta_{(\mu \nu} \eta_{\rho \sigma} \Big[p_{\a} p_{\b)} \, J(d+8; 1, 3, 1 ) \notag \\[8pt] 
  & \qquad - p_{\a} q_{\b)} \, J(d+8; 1, 2, 2 ) +  q_{\a} q_{\b)} \, J(d+8; 1, 1, 3 ) \Big] , \notag \\[5pt] 
  &\quad + 180\, \eta_{(\mu \nu} \Big[ p_{\rho} p_{\sigma} p_{\a} p_{\b)} \, J(d+10; 1, 5, 1 ) -  p_{\rho} p_{\sigma} p_{\a} q_{\b)} \, J(d+10; 1, 4, 2 ) \notag \\[8pt] 
    &\qquad + p_{\rho} p_{\sigma} q_{\a} q_{\b)} \, J(d+10; 1, 3, 3 ) -  p_{\rho} q_{\sigma} q_{\a} q_{\b)} \, J(d+10; 1, 2, 4 ) \notag \\[5pt] 
      &\qquad + q_{\rho} q_{\sigma} q_{\a} q_{\b)} \, J(d+10; 1, 1, 5 )\Big] + 720  \Big[ p_{\mu}  p_{\nu} p_{\rho} p_{\sigma} p_{\a} p_{\b}  \, J(d+12; 1, 7, 1 ) \notag \\[5pt]
    &\qquad - p_{(\mu}  p_{\nu} p_{\rho} p_{\sigma} p_{\a} q_{\b)}  \, J(d+12; 1, 6, 2 ) + p_{(\mu}  p_{\nu} p_{\rho} p_{\sigma} q_{\a} q_{\b)}  \, J(d+12; 1, 5, 3 ) \notag \\[5pt]
        &\qquad - p_{(\mu}  p_{\nu} p_{\rho} q_{\sigma} q_{\a} q_{\b)}  \, J(d+12; 1, 4, 4 ) + p_{(\mu}  p_{\nu} q_{\rho} q_{\sigma} q_{\a} q_{\b)}  \, J(d+12; 1, 3, 5 ) \notag \\[5pt]
                &\qquad - p_{(\mu}  q_{\nu} p_{\rho} q_{\sigma} q_{\a} q_{\b)}  \, J(d+12; 1, 2, 6 ) + q_{\mu}  q_{\nu} q_{\rho} q_{\sigma} q_{\a} q_{\b}  \, J(d+12; 1, 1, 7 ) \Big],
 \end{align}
where the integrals on the l.h.s.\  are all in $d$ dimensions, whereas the dimension
varies on the r.h.s.\ .
Here, as elsewhere in this paper,  all symmetrisations are with strength one. The scalar 
integrals on the r.h.s.\  are now of type (\ref{Jdnu}), but they still involve different dimensions 
$D\,=\,d\,,\, d+2\, ,...$ and different exponents $m_1,m_2,m_3$.  
To further simplify the above expressions we exploit the basic result  \cite{Davyd1,Davyd2} that for integer $m_i$ all integrals of the form (\ref{Jdnu}) 
can be expressed in terms of $J(d;1,1,1)$ and boundary integrals of the type (\ref{Ipd}),  (\ref{Ipd1}).  

The first part in this reduction procedure is to decrease the values $m_1,m_2,m_3$ 
in integer steps while leaving the dimension unchanged; this is done by noting that
 \begin{equation} \label{idred}
  \int \frac{d^dk}{\pi^{d/2}} \frac{\partial}{\partial k_{\mu}} \left\{ \frac{k_{\mu}}{k^{2 m_1} (k -p)^{2 m_2} (k + q)^{2 m_3} }\right\} =0,
 \end{equation}
which gives a relation between $J(D; m_1, m_2, m_3 )$ with 
$\sum m_i =m$ and $J(D; m_1, m_2, m_3 )$ with $\sum m_i =m -1.$ Two more relations can be found by changing the numerator in the integrand in \eqref{idred} to $k_{\mu} -p_{\mu}$ and $k_{\mu} +q_{\mu}$. These three equations can be solved \cite{Davyd2} to obtain 
 \begin{align}
  J({m_1, m_2, m_3+1}) &= \frac{1}{2 m_3 (p+q)^2 q^2} \Bigg( \Big( (2 m_1 + m_2 + m_3 - d) \, (p+q)^2  \notag \\[3mm]
  & \hspace{6mm} + (m_1 + 2 m_2 + m_3- d) \, q^2 - ( m_1 + m_2 + 2 m_3 - d) \, p^2 \Big) J({m_1, m_2, m_3 })  \notag \\[3mm]
 & \quad + m_2 \, (p+q)^2 J({ m_1-1, m_2 + 1, m_3 }) + m_3 \, (p+q)^2 J({ m_1-1, m_2, m_3 + 1})  \notag \\[3mm]
  & \quad +  m_1 \, q^2 J({ m_1 + 1, m_2 - 1, m_3}) + m_3 \, q^2 J({ m_1, m_2 - 1, m_3 + 1})  \notag \\
  & \quad -  m_1 \, p^2 J({ m_1 +1 , m_2, m_3 -1}) - m_2 \, p^2 J({ m_1, m_2 + 1, m_3 -1}) \Bigg), \label{redid1}
 \end{align}
 
 \begin{align}
  J({m_1 + 1, m_2, m_3}) &= \frac{1}{2 m_1 p^2 q^2} \Bigg( \Big( ( m_1 + 2 m_2 + m_3 - d) \, q^2 + (m_1 + m_2 + 2 m_3- d) \, p^2 \notag \\
  & \hspace{33mm}  - ( 2 m_1 + m_2 + 2 m_3 - d) \, (p+q)^2 \Big) J({m_1, m_2, m_3 })  \notag \\[3mm]
 & \quad + m_2 \, p^2 J({ m_1, m_2 + 1, m_3 -1 }) + m_1 \, p^2 J({ m_1+1, m_2, m_3 - 1})  \notag \\[3mm]
  & \quad +  m_3 \, q^2 J({ m_1, m_2 - 1, m_3 +1}) + m_1 \, q^2 J({ m_1 + 1, m_2 - 1, m_3})  \notag \\[3mm]
  & \quad -  m_2 \, (p+q)^2 J({ m_1 -1 , m_2 +1, m_3}) \notag \\
 & \quad - m_3 \, (p+q)^2 J({ m_1 -1, m_2, m_3 +1}) \Bigg), \label{redid2}
 \end{align}
 
  \begin{align}
  J({m_1, m_2+1, m_3}) &= \frac{1}{2 m_2 (p+q)^2 p^2} \Bigg( \Big( (2 m_1 + m_2 + m_3 - d) \, (p+q)^2  \notag \\
  & \hspace{6mm} + (m_1 + m_2 +2 m_3- d) \, p^2 - ( m_1 + 2 m_2 + m_3 - d) \, q^2 \Big) J({m_1, m_2, m_3 })  \notag \\[3mm]
 & \quad + m_3 \, (p+q)^2 J({ m_1-1, m_2 , m_3+1 }) + m_2 \, (p+q)^2 J({ m_1-1, m_2+1, m_3})  \notag \\[3mm]
  & \quad +  m_1 \, p^2 J({ m_1 + 1, m_2 , m_3 -1}) + m_2 \, p^2 J({ m_1, m_2 + 1, m_3 - 1})  \notag \\
  & \quad -  m_1 \, q^2 J({ m_1 +1 , m_2-1, m_3}) - m_3 \, q^2 J({ m_1, m_2 - 1, m_3 +1}) \Bigg). \label{redid3}
 \end{align}
We repeat that the dimension $D$ is the {\em same} in all these integrals, whence
$$J({m_1, m_2, m_3 }) \equiv J(D; m_1, m_2, m_3 )$$ with the same $D$ 
in the three equations above. 

Having reduced the integrals $J_{\mu_1 \cdots \mu_M}$ of the form given 
in equation \eqref{tensorint} to a sum of scalar integrals $J(D; {1, 1, 1})$ and boundary 
integrals, where $D\,=\, d\,,\, d+2\, , \cdots, \,d+ 2M$ we next require a further identity which 
lowers the values of $D$ by relating $J(D+2; {1, 1, 1})$ to $J(D; {1, 1, 1})$ so that 
finally all integrals can be reduced to $J(d; {1, 1, 1})$, where now $d = 4 - 2\epsilon$. 
The relevant identity is found 
by contracting the indices in equation \eqref{3int2}, whereupon the l.h.s.\ of (\ref{3int2}) 
reduces to a boundary integral, and we get
\begin{align}
  J(d;0,1,1) &= \frac{d}{2} \, J(d+2; 1, 1, 1 ) + 2 \Big[ p^2 \, J(d+4; 1, 3, 1 ) \notag \\[5pt]
  & \hspace{20mm} - \pq \, J(d+4; 1, 2, 2 ) + q^2 \, J(d+4; 1, 1, 3 ) \Big].
\end{align}
Then, using the reduction formulae \eqref{redid1}--\eqref{redid3}, we express 
$J(d+4; 1, 3, 1 ) , J(d+4; 1, 2, 2 )$ and $ J(d+4; 1, 1, 3)$ in terms of $J(d+4; {1, 1, 1})$ and boundary integrals. Substituting, these expression in the equation above, and 
replacing $d \rightarrow d-2$, we obtain 
\begin{align}
 & 2(d-2) \left( \pq^2 - p^2 q^2\right)J(d+2;1,1,1) - p^2 q^2 (p+q)^2 J(d;1,1,1) \notag \\[3mm]
    = \, & \pq  (p+q)^2 \, J(d; 0, 1, 1 ) - p^2 (\pq  +q^2) J(d; 1, 1, 0 ) - q^2 ( p^2 + \pq) J(d; 1, 0, 1 ),
\end{align}
or
\begin{align}\label{Jd+2tod}
J(d+2;1,1,1\,|\,p,q) &= \frac1{2(d-2)} \frac1{\pq^2 - p^2 q^2} \,\bigg[
p^2 q^2 (p+q)^2 J(d;1,1,1\,|\, p,q)  \, + \nn\\[2mm]
&\!\!\!\!\!\!\!\!\!\!\!\!\!\!\!\!\!\!\!\!\!\!\!\!\!\!\!\!\!\!\!\!\!\!\!\!\!\!\!\!\!\!\!\!\!\!\!\!\!\!\!\!
 + \, \pq  (p+q)^2 \, I(d \,|\,p+q)  \,-\, p^2 (\pq  +q^2) I(d\,|\, p) \,-\, q^2 ( p^2 + \pq) I(d \,|\, q ) \bigg].
\end{align}
where we have now substituted the boundary integrals from the appendix \ref{app:int}.
This formula seems to be a new result: it allows us to reduce any given integral of
type (\ref{tensorint}) for even $d$ to sums involving the convergent integral $J(4;1,1,1)$ and
various boundary integrals which contain all the UV divergences as $d\rightarrow 4$. 
Using the formula \eqref{I(p)} from appendix \ref{app:int}, the latter can be exhibited explicitly:
\begin{equation}
J(d+2;1,1,1) = - i\, \frac{d-1}{d-2} \cdot \frac{\Gamma(1- \frac{d}{2}) \Gamma(\frac{d}{2})^2}{\Gamma(d)}
  \, + \, \mbox{finite terms}. \label{Jd2111}
\end{equation}
The factor $(p^2q^2 - \pq^2)^{-1}$ is cancelled, whence the UV divergence 
does not depend on the external momenta, as expected. Furthermore, formula \eqref{Jd+2tod}
makes the kinematical singularities completely explicit. We note that the above formulae, \eqref{Jd+2tod} and \eqref{Jd2111} cannot
be used for $d=2$ because of IR singularities and the factor $(d-2)^{-1}$.

Using equations  \eqref{redid1}--\eqref{redid3} and \eqref{Jd+2tod}, the integrals $J_{\mu_1 \cdots \mu_M}$ can thus be reduced to boundary integrals and $J(d,1,1,1)$. Since $J(4,1,1,1)$ is finite, the $1/\epsilon$ poles in $J_{\mu_1 \cdots \mu_M}$ can easily be found by expanding $d= 4 -2\epsilon$ and using the result for the poles of the boundary integral, \eqref{I(p)}. The $1/\epsilon$ expansion of $J_{\mu_1 \cdots \mu_M}$ up to $M=6$ is:
\begin{align}
  J_{\mu}(d\,|\,p,q) &= O(1), \label{3int1}\\[3pt]
  J_{\mu \nu}(d\,|\,p,q) &= \frac{i}{4 \epsilon} \eta_{\mu \nu} + O(1),  \\[3pt]
  J_{\mu \nu \rho}(d\,|\,p,q) &= \frac{i}{4 \epsilon}\eta_{(\mu \nu}  \left( p_{\rho)} - q_{\rho)} \right) + O(1), \\[3pt] 
       J_{\mu \nu \rho \sigma}(d\,|\,p,q) &= - \frac{i}{32 \epsilon} \left( p^2 + q^2 + (p+q)^2 \right) \eta_{(\mu \nu} \eta_{\rho \sigma)} + \frac{i}{4 \epsilon} \, \eta_{(\mu \nu} \left( p_{\rho} p_{\sigma)} - p_{\rho} q_{\sigma)}  + q_{\rho} q_{\sigma)} \right) + O(1), \\[3pt]  
       J_{\mu \nu \rho \sigma \alpha}(d\,|\,p,q) &= - \frac{i}{32 \epsilon}  \eta_{(\mu \nu} \eta_{\rho \sigma} \left[ \left( 2 p^2 + q^2 + 2 (p+q)^2 \right) p_{\a)}  - \left( p^2 + 2 q^2 + 2 (p+q)^2 \right) q_{\a)} \right] \notag \\[3pt] 
       &\quad + \frac{i}{4 \epsilon}\,  \eta_{(\mu \nu} \left( p_{\rho} p_{\sigma} p_{\a)} - p_{\rho} p_{\sigma} q_{\a)} +  p_{\rho} q_{\sigma} q_{\a)} - q_{\rho} q_{\sigma} q_{\a)} \right) + O(1), \\[3pt] 
       J_{\mu \nu \rho \sigma \a \b}(d\,|\,p,q) &= \frac{i}{192 \epsilon} \left[ \left( p^2 + q^2 + (p+q)^2 \right)^2 - (p + q)^2(p^2 + q^2) - p^2 q^2 \right] \eta_{(\mu \nu} \eta_{\rho \sigma} \eta_{\a \b)}  \notag \\[3pt] 
       & \quad  - \frac{i}{32 \epsilon} \eta_{(\mu \nu} \eta_{\rho \sigma} \left[ \left( 3 p^2 + q^2 + 3 (p+q)^2 \right) p_{\a} p_{\b)}  
  - 2 \left( p^2 + q^2 + 2 (p+q)^2 \right) p_{\a} q_{\b)}  \right. \notag \\[5pt]
  & \quad \left.  + \left( p^2 + 3 q^2+ 3 (p+q)^2 \right) q_{\a} q_{\b)}  \right] + \frac{i}{4 \epsilon} \, \eta_{(\mu \nu} \left(p_{\rho} p_{\sigma} p_{\a} p_{\b)}-  p_{\rho} p_{\sigma} p_{\a} q_{\b)} \right. \notag \\[5pt] 
  &\quad\qquad\qquad 
  \left.  + \, p_{\rho} p_{\sigma} q_{\a} q_{\b)}  -  p_{\rho} q_{\sigma} q_{\a} q_{\b)}  + q_{\rho} q_{\sigma} q_{\a} q_{\b)} \right) + O(1). \label{3int6}
 \end{align}
These coefficients and polynomials in the external momenta are what we need for the
evaluation of the conformal anomaly.

If one is just interested in the divergent parts, this result can also be arrived
at without invoking the full machinery of $n$-point
loop integrals and in a much simpler way as follows: First of all, one notes that the divergence must be
polynomial in the external momenta $p$ and $q$. Secondly the resulting
polynomial must be symmetric under interchange of $p$ and $-q$. Thirdly, by
shifting the integration variable as $k \rightarrow -k+p$ one obtains a relation 
constraining the polynomials by replacing the external momenta $(p,q)$ by $(p, -p-q)$. 
When applying this trick to the above integrals, one first notes that the integrals $J$ and
$J_\mu$ are convergent, whence the first divergence arises in $J_{\mu\nu}$; the 
latter divergence is proportional to $\eta_{\mu\nu}$ and can thus be 
extracted by contracting with $\eta_{\mu\nu}$, thereby cancelling one propagator and
reducing the determination of the pole term to that of a 2-point integral.
Likewise the divergence in $J_{\mu\nu\rho}$ can only appear in the term
linear in the external momenta, which by symmetry must appear in the 
combination $(p-q)_\mu$; again the result can be read off from the corresponding
2-point integral after contraction, and so on for the integrals with more momenta in the numerator.

It is easy to see that this procedure can also be applied inductively to $n$-point
integrals for $n>3$ by successively reducing them to $(n-1)$-loop integrals, {\it etc.}
In other words, the determination of the pole parts at any order in $h_{\mu\nu}$
does not require the actual evaluation of $n$-point integrals. However, this shortcut
may no longer be available for classically non-conformal theories where there could 
arise extra non-local contributions.

\section{The conformal anomaly at $\cO(h^2)$} \label{sec:anom}

The anomaly is given by the trace of $\langle T_{\mu \nu}(x) \rangle$ \emph{after} regularisation. If we calculate the trace before finding the regularised expression, the trace vanishes by the Ward identities as the Dirac action is scale-invariant in all dimensions. At second order in the external graviton, the anomaly is given by 
\begin{align}
 g^{\mu \nu} \langle T_{\mu \nu}(x) \rangle\big|_{\cO(h^2)}\, &= \eta^{\mu \nu} \langle T_{\mu \nu} \rangle|_{\mathcal{O}(h^2)} - h^{\mu \nu}(x)  \langle T_{\mu \nu} \rangle|_{\mathcal{O}(h)} \notag \\[9pt]
 & = \int d^d y  \,  d^dz \int\frac{d^dp}{(2 \pi)^d}  \frac{d^dq}{(2 \pi)^d} e^{-i p\cdot (x - y) -i q\cdot (z - y)} h^{\rho \sigma}(y)    \notag \\[5pt]
  & \hspace{5mm} \times \Biggl\{ \eta^{\mu \nu}\Bigg( T^{(1)}_{\mu\nu \rho \sigma \a \b} (p, q) \,  h^{\a \b}(z)+ T^{(2)}_{\mu\nu \rho \sigma \a \b} (p,q) \,  h^{\a \b}(z) + T^{(3)}_{\mu\nu \rho \sigma \a \b} (p) \,  h^{\a \b}(x) \Bigg) \notag \\[5pt]
 & \hspace{35mm} - T_{\mu \nu \rho \sigma }(p)  h^{\mu \nu}(x) \Biggr\}.  \label{trace}
\end{align}
Using equations \eqref{T31}, \eqref{T32} and \eqref{T33}, and rewriting
\begin{align*}
&\int d^d y  \int\frac{d^dp}{(2 \pi)^d} e^{-i p\cdot (x - y)}  h^{\mu \nu}(x) h^{\rho \sigma}(y)  T_{\mu \nu \rho \sigma }(p)  \\[7pt]
& \hspace{35mm}= \int d^d y  \,  d^dz \int\frac{d^dp}{(2 \pi)^d}  \frac{d^dq}{(2 \pi)^d} e^{-i p\cdot (x - y) -i q\cdot (z - y)} h^{\rho \sigma}(y) h^{\mu \nu}(z) T_{\mu \nu \rho \sigma }(p+q),
\end{align*}
\begin{align*}
 g^{\mu \nu} \langle T_{\mu \nu}(x) \rangle \,  & =  \frac{1}{4} \, \eta^{\mu \nu} \int d^d y  \,  d^dz \int\frac{d^dp}{(2 \pi)^d}  \frac{d^dq}{(2 \pi)^d} e^{-i p\cdot (x - y) -i q\cdot (z - y)} h^{\rho \sigma}(y)   h^{\a \b}(z)  \notag \\[5pt]
  & \hspace{25mm} \times  \Biggl(\frac18 \, T_{\mu \nu \rho \sigma \a \b} + \eta_{\mu \nu} \, T_{\rho \sigma \a \b }(p+q) - \eta_{\mu \nu} \, T_{\rho \sigma \a \b }(q) - \eta_{\rho \sigma}  \, T_{\mu\nu\a\b}(q) \notag \\[5pt]
 & \hspace{33mm}  -  \eta_{\a\b} T_{\mu\nu\rho\sigma}(p+q)  - 3\,  \eta_{\a \rho} \, T_{\mu\nu \sigma \b }(p) + 2 \, \eta_{\a \mu} \, T_{\nu \b \rho \sigma  }(p+q) \Biggr). 
\end{align*}
By redefining the integration variables $p$ and $q$, we can write the above expression in a more symmetric way,
\begin{align}
 g^{\mu \nu} \langle T_{\mu \nu}(x) \rangle \,  & =  \frac{1}{4} \, \eta^{\mu \nu} \int d^d y  \,  d^dz \int\frac{d^dp}{(2 \pi)^d}  \frac{d^dq}{(2 \pi)^d} e^{-i p\cdot (x - y) -i q\cdot (x- z)} h^{\rho \sigma}(y)   h^{\a \b}(z)  \notag \\[5pt]
  & \hspace{25mm} \times  \Biggl(\frac18 \, \hat{T}_{\mu \nu \rho \sigma \a \b} - \, \eta_{\a \mu} \, T_{\rho \sigma\nu \b   }(p) - \, \eta_{\rho \mu} \, T_{\nu \sigma \a \b}(q) \notag \\[5pt] 
 & \hspace{59mm} - 2 \, \eta_{\a\b} T_{\mu\nu\rho\sigma}(p)  - 3\,  \eta_{\a \rho} \, T_{\mu\nu \sigma \b }(p+q)  \Biggr),  \label{trace2}
\end{align}
where we have exploited the symmetry under $ p \leftrightarrow q$ and $ \rho \sigma \leftrightarrow \a \b$, to simplify the integrand and where $\hat{T}_{\mu \nu \rho\sigma \a \b}$ is the expression by letting $p \rightarrow p+q$ and $q \rightarrow -q$, \emph{viz}.\ 
\begin{equation}
\hat{T}_{\mu \nu \rho \sigma \a \b} \equiv i \int \frac{d^dk}{(2 \pi)^d} \, \textrm{tr}\biggl\{ \frac{k\!\!\!/}{k^2}  (2k + q )_{(\a} \gamma_{\b)}  \frac{k\!\!\!/ + q\!\!\!/}{(k+q)^2} (2 k - p + q)_{(\mu} \gamma_{\nu)}  \frac{k\!\!\!/ - p\!\!\!/ }{(k-p)^2} (2 k- p)_{(\rho} \gamma_{\sigma)}
\biggr\}.
\end{equation}

The trace of the expectation in $d$-dimensions should be zero as the Dirac action is Weyl-invariant in all dimensions. The anomaly appears because the expectation value of the regularised 4-dimensional stress tensor is evaluated in $d=4-2 \epsilon$ dimensions, which gives rise to a non-zero 4-dimensional trace. As a consistency check we show that the $d$-dimensional trace of the expression of the r.h.s.\  of equation \eqref{trace2} vanishes. 

First consider,
\begin{align}
  \eta^{(d) \, \mu \nu} \, \hat{T}_{\mu \nu \rho \sigma \a \b}
& = i \,  \int \frac{d^dk}{(2 \pi)^d}  \textrm{tr}\biggl\{ \ \frac{k\!\!\!/ }{k^2}  (2 k - p )_{(\rho} \gamma_{\sigma)} 
  \frac{(k\!\!\!/ - p\!\!\!/ )}{(k-p)^2} (2 k  + q )_{(\a} \gamma_{\b)}    \notag  \\[8pt]
 &  \hspace{30mm} + \frac{k\!\!\!/}{k^2}   (2 k+ p )_{(\rho} \gamma_{\sigma)}  
  \frac{(k\!\!\!/ -q\!\!\!/ )}{(k-q)^2} (2 k - q )_{(\a} \gamma_{\b)} \biggr\}, \notag \\[5pt]
&=  \,  8 \big(T_{\rho \sigma \a \b}(p) + T_{\rho \sigma \a \b}(q)\big), \label{Thattrace}
\end{align}
where we have used an identity similar to identity \eqref{ident} and reparametrised the  variable of integration $k$ in the first equality, and equation \eqref{niceid} and the definition of the two-point function integral, \eqref{2pt}, in the second equality.
Therefore, substituting into equation \eqref{trace2} and using the Weyl-invariance of the 2-point function in $d$-dimensions, \eqref{1trace} ,
\begin{align}
\big\langle  g^{\mu \nu}  T_{\mu \nu}(x) \big\rangle\big|_{d-\textrm{dim}}    =  0. \label{ddimtrace}
\end{align}

Therefore, from equation \eqref{trace2}, the anomaly at second order in $h$ is given by
\begin{align} \label{panom}
 \mathcal{A}(x)\big|_{\mathcal{O}(h^2)}& =    \frac{1}{4}  \int d^d y  \,  d^dz \int\frac{d^dp}{(2 \pi)^d}  \frac{d^dq}{(2 \pi)^d} e^{-i p\cdot (x - y) -i q\cdot (z - y)} h_{\rho \sigma}(y)   h_{\a \b}(z)  \notag \\[5pt]
   & \hspace{15mm} \times  \Biggl(\frac18 \, \eta^{(4) \, \mu \nu} \hat{T}_{\mu \nu \rho \sigma \a \b} - \,  T_{\rho \sigma \a \b }(p)  - \, T_{\rho \sigma \a \b }(q)  \notag \\[5pt] 
 & \hspace{30mm} - 2 \,  \eta^{(4)\eta_{\a\b} \, \mu \nu} \, T_{\mu  \nu \rho \sigma }(p) - 3 \,\eta^{(4) \, \mu \nu} \eta_{ \rho \a } T_{\mu\nu\sigma\b}(p+q)   \Biggr), 
\end{align}
where $ \eta^{(4) \, \mu \nu} \hat{T}_{\mu\nu \a \b \rho \sigma}$ is the 4-dimensional trace of the regularised 3-point function in momentum space.

The 4-dimensional trace of the regularised 2-point function integral is already known and given in equation \eqref{2pttrace}. Therefore, it remains to consider the terms on the second line of the r.h.s.\  of equation \eqref{panom}. We write,
\begin{equation}
 \frac18 \hat{T}_{\mu \nu \rho \sigma \a \b} - \, \eta_{\a \mu} \, T_{\rho \sigma\nu \b   }(p) - \, \eta_{\rho \mu} \, T_{\nu \sigma \a \b}(q)  = \frac{A_{\mu \nu \rho \sigma \a \b}(p,q)}{2 \epsilon} + B_{\mu \nu \rho \sigma \a \b}(p,q) + \mathcal{O}(\epsilon). \label{partofint}
\end{equation}
The terms on the l.h.s.\  are regularised integrals in $d$-dimensions and we denote the pole terms in the expression by $A_{\mu \nu \rho \sigma \a \b}$ and the finite terms by $B_{\mu \nu \rho \sigma \a \b}.$ We are interested in the 4-dimensional trace of the expression on the l.h.s.\ , which gives the terms on the second line of the r.h.s.\  of equation \eqref{panom}. Namely, we are interested in 
\begin{equation}
\eta^{(4) \, \mu \nu} B_{\mu \nu \rho \sigma \a \b} =  \frac18 \, \eta^{(4) \, \mu \nu} \hat{T}_{\mu \nu \rho \sigma \a \b} - \, T_{\rho \sigma \a \b }(p)  - \, T_{\rho \sigma \a \b }(q). \label{B4trace}
\end{equation}
Note that the 4-dimensional trace of $A_{\mu \nu \rho \sigma \a \b}$ necessarily vanishes, since the anomaly is finite. 

The tensor $A_{\mu \nu \rho \sigma \a \b}$ is local in the momenta $p$ and $q$ and can be found using equations \eqref{3int1}--\eqref{3int6}. Meanwhile, the tensor $ B_{\mu \nu \rho \sigma \a \b}$  is given by the terms labelled $\mathcal{O}(1)$ in equations \eqref{3int1}--\eqref{3int6} and is in 
general non-local in the momenta.
The 4-dimensional trace of $B_{\mu \nu \rho \sigma \a \b}$ can nevertheless be found from $A_{\mu \nu \rho \sigma \a \b}$ by taking a trace in $D$ dimensions, where $D$ is arbitrary (but remember that
the 2-point and 3-point functions above are computed in $d$-dimensions, so $D$ is 
just an auxiliary variable here).

From equation \eqref{Thattrace}, we know that
\begin{align}
\eta^{(D) \, \mu \nu} \left( \frac 18 \hat{T}_{\mu \nu \rho \sigma \a \b} - \, \eta_{\a \mu} \, T_{\rho \sigma\nu \b   }(p) - \, \eta_{\rho \mu} \, T_{\nu \sigma \a \b}(q)   \right) = (D- d) \left( \frac{C_{\rho \sigma \a \b}}{2\epsilon} + D_{\rho \sigma \a \b} + \mathcal{O}(1) \right),  \label{partofint2}
\end{align}
where $C_{\rho \sigma \a \b}$ and $D_{\rho \sigma \a \b}$ are tensorial functions of the momenta. 
Substituting equation \eqref{partofint} on the l.h.s.\  of equation \eqref{partofint2} and expanding the r.h.s.\  in $\epsilon$, we find
\begin{align}
 \eta^{(D) \, \mu \nu} A_{\mu \nu \rho \sigma \a \b} &= (D- 4) C_{\rho \sigma \a \b},  \label{Atrace} \\
 \eta^{(D) \, \mu \nu} B_{\mu \nu \rho \sigma \a \b} &= C_{\rho \sigma \a \b} + (D- 4) D_{\rho \sigma \a \b},  \label{Btrace}
\end{align}
at order $1/\epsilon$ and order 1. 
Letting $D=4$ in equation \eqref{Btrace}, and using \eqref{B4trace}, we find that 
\begin{equation}
 C_{\rho \sigma \a \b} = \frac18 \, \eta^{(4) \, \mu \nu} \hat{T}_{\mu \nu \rho \sigma \a \b} - \, T_{\rho \sigma \a \b }(p)  - \, T_{\rho \sigma \a \b }(q). \label{Cdef}
\end{equation}
However,  $C_{\rho \sigma \a \b}$ can also be found by taking the $D$-dimensional trace of $A_{\mu \nu \rho \sigma \a \b}$, \eqref{Atrace}. 

After a lengthy calculation (that  involves collecting several hundred terms!)
we determine $A_{\mu \nu \rho \sigma \a \b}$, defined in equation \eqref{partofint}, and identify  $C_{\rho \sigma \a \b}$ by taking an arbitrary $D$-dimensional trace, \eqref{Atrace}. This gives, \eqref{Cdef}, an expression for the terms on the second line of the r.h.s.\ of equation \eqref{panom} and, as we have already mentioned, the other terms on the r.h.s.\  of equation \eqref{panom} are given by equation \eqref{2pttrace}. The final result is
\begin{align}
 \mathcal{A}|_{\mathcal{O}(h^2)} &= - \frac{1}{ 360 \,  (4 \pi)^{2}} \int d^4 y  \,  d^4z \int\frac{d^4p}{(2 \pi)^4}  \frac{d^4q}{(2 \pi)^4} e^{-i p\cdot (x - y) - i q\cdot (x - z)} h^{\rho \sigma}(y)   h^{\a \b}(z) \notag \\[3mm]
     & \qquad \times\Bigg\{  ( (p\cdot q)^2 + p^2 ( 12 \, p\cdot q + 5 q^2))\, 
     \eta_{\rho \sigma} \eta_{\a \b} \notag \\[3mm] & \qquad 
     - (12 \, (p^2)^2 + 25\, (p\cdot q)^2 + 14 \, p^2\, (3 \, p\cdot q +  q^2)) \, 
     \eta_{\a (\rho} \eta_{\sigma) \b}  \notag \\[3mm] & \qquad
 - 8 \,(3\,  p^2 + 3 \,  p\cdot q + 2 \, q^2) \,  \eta_{\rho \sigma} p_{\a} p_{\b} - 2 \, \,(3 \,  p\cdot q + 5 \, q^2)\, 
 \eta_{\a \b} p_{\rho} p_{\sigma} \notag \\[3mm] &  \qquad
  -  4\, (3 \, p^2 +4 \, p \cdot q +6 \, q^2)) \, \eta_{\a \b} p_{(\rho} q_{\sigma)}  
  + 8\, (3 \, p^2 + 6\, p \cdot q + 4 \, q^2)  \, p_{(\rho} \eta_{\sigma) (\a}  p_{\b)}\notag \\[3mm] & \qquad 
  +  4 \, (6 \, p^2 + 5 \, p \cdot q)  \,  p_{(\rho} \eta_{\sigma) (\a}  q_{\b)} + 2\,  (6 \, p^2 + 13 \,  p\cdot q)  \, q_{(\rho} \eta_{\sigma) (\a}  p_{\b)}\notag \\[3mm]  & \qquad
 +12 \,p_{\rho} p_{\sigma} p_{\a} p_{\b} + 12 \,p_{\rho} p_{\sigma} p_{(\a} q_{\b)}  
+5  \,p_{\rho} p_{\sigma} q_{\a} q_{\b}  
-4 \, p_{(\rho} q_{\sigma)} p_{(\a} q_{\b)} -7  \, q_{\rho} q_{\sigma} p_{\a} p_{\b} \Bigg\}.
\end{align}
The expression is, in particular, polynomial in $p$ and $q$ -- the dependence on inverse powers
or logarithms of the external momenta, which are in higher order terms in $\epsilon$, has 
dropped out, hence the anomaly is local in $x$-space, as expected.

Note that when comparing with the anomalies the terms quadratic in curvature must all have
the structure $\partial\partial h \partial\partial h$, which in Fourier space is equivalent to having two $p$ 
and two $q$ in each term, whereas all other terms with a different distribution of derivatives must originate from $\Box R$. Therefore, we can use the term proportional to $ \,p_{\rho} p_{\sigma} p_{\a} p_{\b}$ (see equation \eqref{br2s}), for example, to fix the coefficient of $\Box R$ , 
\begin{equation}
  \mathcal{A}\big|_{\mathcal{O}(h^2)} =  \frac{1}{30(4 \pi)^{2}} \Box R \big|_{\cO(h^2)}  + \dots.
\end{equation}
Furthermore, from equation \eqref{riem2s}--\eqref{r2s}, we note that $q_{\rho} q_{\sigma} p_{\a} p_{\b}$, $ p_{(\rho} q_{\sigma)} p_{(\a} q_{\b)}$, $p_{\rho} p_{\sigma} q_{\a} q_{\b} $ only appear in Riemann-squared, Ricci-squared and scalar-squared, respectively. Hence terms containing these expressions can be used to fix the coefficient of all the terms in the anomaly. Altogether we have thus shown that
\begin{align}
  \mathcal{A}|_{\mathcal{O}(h^2)} &= \bigg[
  \frac{7}{360(4 \pi)^{2}} \textup{Riem}^2 + \frac{1}{45(4 \pi)^{2}} \textup{Ric}^2 - \frac{1}{72(4 \pi)^{2}} R^2 + \frac{1}{30(4 \pi)^{2}} \Box R \bigg]_{\cO(h^2)}  \notag \\[5pt]
  &=  \bigg[  \frac{1}{20(4 \pi)^{2}}  \, C^{\mu\nu\rho\sigma}C_{\mu\nu\rho\sigma} - \frac{11}{360(4 \pi)^{2}} \textup{E}_{4} + \frac{1}{30(4 \pi)^{2}} \Box R\bigg]_{\cO(h^2)} .
\end{align}
Note that the coefficient of $\Box R$ at second order in $h$ matches the coefficient at first order, \eqref{anom1oder}, as it must do for consistency. Furthermore, this explicit calculation confirms
the relation (\ref{bc}), and agrees with the values for $a,b,c$ in the literature.

\section{Outlook}
In this paper we have given a new and direct derivation of the spin-$\frac12$ anomaly,
along the lines of the textbook derivation of the axial anomaly.  Although at this point
the calculation merely confirms a known result, our derivation based on standard Feynman 
diagram techniques has brought out several subtleties, and we expect similar subtleties
for a rederivation of the (again known) results for $s=0,1$.

However, as we already said in the introduction, the present work should be regarded as 
only preparatory for what we are really after, namely a proper computation of and a better 
understanding of the conformal anomaly in {\em non-conformal} theories, where the 
anomaly can be {\em defined} by
\beq
\cA \, := \, g^{\mu\nu} \langle T_{\mu\nu} \rangle \, - \, \langle g^{\mu\nu} T_{\mu\nu} \rangle 
\eeq
and where the second term subtracts the terms due to the classical violation of Weyl
invariance. Most significantly we will be interested in the cases $s=\frac32$ and $s=2$, 
where there remain several issues (dependence of $a$ and $c$ coefficients on gauge
choices for external gravitons, appearance of $R^2$ contributions for non-conformal 
theories, etc.) that remain open even after many years.  Future directions are thus:
\bit
\item A computation of conformal anomaly for $s=\frac32$ along the lines of this paper.
\item Understanding the appearance of $R^2$ and possible non-local contributions
         that may be required to satisfy WZ consistency condition.
\item Understanding the dependence of $a$ and $c$ coefficients on the choice of
         gauge for metric fluctuation $h_{\mu\nu}$. Such a gauge dependence
         should not exist, as the anomaly coefficients should be {\em gauge invariant}
         with the (natural) assumption of unbroken general covariance.
\item Understanding the appearance of negative anomaly coefficients for $s=\frac32$,
         which is in apparent conflict with positivity theorems. However, the latter 
          rely on unitarity (positive definite) Hilbert space, and the existence
          of a gauge invariant stress tensor, whereas both these assumptions are
          violated for $s\geq \frac32$.
\eit

\vspace{8mm}\noindent
{\bf Acknowledgments:} We would like to thank T.~Bautista, L.~Casarin, M.~Duff, K.~Meissner and
A.~Schwimmer for discussions related to this work, and L.~Bonora for correspondence and
explanations concerning refs. \cite{Bonora1,Bonora2}.

\newpage
\appendix

\section{Expansions} \label{app:exp}

In this appendix we collect all necessary formulae for the 
expansion of curvature-squared quantities to second order in $h$:
\begin{align}
g_{\mu \nu} &= \eta_{\mu \nu} + h_{\mu \nu}, \hspace{34.5mm} g^{\mu \nu} = \eta^{\mu \nu} - h^{\mu \nu} + h^{\mu \rho} h^{\nu \sigma} \eta_{\rho \sigma}, \\[3mm]
e_{\mu}{}^a &= \delta_{\mu}^{a} + \frac12 h_{\mu}{}^{a} - \frac18  h_{\mu \nu} h^{\nu a}, \hspace{12.5mm} e^{\mu}{}_{a} = \delta^{\mu}_{a} - \frac12 h^{\mu}{}_{a} + \frac{3}{8} h^{\mu \nu} h_{\nu a}, \\[3mm]
e &= 1+ \frac12 h - \frac14 h^{\mu \nu} h_{\mu \nu} + \frac18 h^2, \qquad e^{-1} = 1 - \frac12 h + \frac14 h^{\mu \nu} h_{\mu \nu} + \frac18 h^2, \\[3mm]
 \omega_{\mu \, a b} &= \partial_{[b} h_{a] \mu} + \frac14 h^{\nu}{}_{[b|} \partial_{\mu} h_{|a] \nu} - \frac12  h^{\nu}{}_{[b|} \left( \partial_{\nu} h_{|a] \mu} - \partial_{a]} h_{\mu \nu} \right), \\[3mm]
 \textrm{Riem}^2 &= \partial^{\mu} \partial^{\nu} h^{\rho \sigma} \left( \partial_{\mu} \partial_{\nu} h_{\rho \sigma} - 2 \, \partial_{\mu} \partial_{\rho} h_{\nu \sigma} + \partial_{\rho} \partial_{\sigma} h_{\mu \nu} \right), \label{riem2}\\[3mm]
  \textrm{Ric}^2 &= \frac{1}{2} \partial_{\mu} \partial^{\rho} h^{\mu \sigma} \left( \partial^{\nu} \partial_{\rho} h_{\nu \sigma} - 2 \,  \partial^2 h_{\rho \sigma}  - 2 \, \partial_{\rho} \partial_{\sigma} h + \partial^{\nu} \partial_{\sigma} h_{\nu \rho} \right) \notag \\[3mm]
  & \quad  + \frac14 \partial^2 h^{\rho \sigma} \left(  \partial^2 h_{\rho \sigma} + 2\, \partial_{\rho} \partial_{\sigma} h \right) + \frac14 \partial^{\rho} \partial^{\sigma} h \, \partial_{\rho} \partial_{\sigma} h , \label{ric2}\\[3mm]
   R^2 &= \partial^{\rho} \partial^{\sigma} h_{\rho \sigma} \, \partial^{\mu} \partial^{\nu} h_{\mu \nu}  - 2 \,  \partial_{\rho} \partial_{\sigma} h_{\rho \sigma} \, \partial^2 h +  \partial^2 h \, \partial^2  h,\label{r2} \\[3mm]
   \Box R &= \partial^2 \partial^{\rho} \partial^{\sigma} h_{\rho \sigma} - \partial^2 \partial^2 h  \\[2mm]
   & \quad + h^{\rho \sigma} \left( 2 \, \partial^2 \partial_{\rho} \partial_{\sigma} h - 2 \, \partial^2 \partial^{\mu} \partial_{\rho} h_{\mu \sigma} +  \partial^2 \partial^2 h_{\rho \sigma} -  \partial_{\rho} \partial_{\sigma} \partial^{\mu} \partial^{\nu} h_{\mu \nu} \right) \notag \\[2mm]
    & \quad + \partial^{\mu} h^{\rho \sigma} \left( 2 \, \partial_{\mu} \partial_{\rho} \partial_{\sigma} h - 4 \,  \partial_{\mu} \partial_{\rho} \partial^{\nu} h_{\nu \sigma} + \frac{7}{2} \, \partial^2 \partial_{\mu} h_{\rho \sigma} -  \partial^2 \partial_{\rho} h_{\sigma \mu} \right) \notag \\[2mm]
     & \quad - \left(\partial_{\mu} h^{\mu \rho} - \frac 12 \partial^{\rho} h \right) \left( 2 \, \partial^2 \partial^{\nu} h_{\rho \nu} - 2 \, \partial_{\rho} \partial^2  h + \partial_{\rho} \partial^{\nu} \partial^{\sigma} h_{\nu \sigma} \right) \notag \\[2mm]
       & \quad + \frac{1}{2} \partial^{\mu} \partial^{\nu} h^{\rho \sigma} \left( 3 \, \partial_{\mu} \partial_{\nu} h_{\rho \sigma} - 2 \, \partial_{\mu} \partial_{\rho}  h_{\nu \sigma} \right) - 2 \, \partial^{\rho} \partial_{\mu} h^{\mu \sigma} \left( \partial_{\rho} \partial^{\nu} h_{\nu \sigma} - \partial_{\rho} \partial_{\sigma}  h + \partial^2 h_{\rho \sigma} \right) \notag \\[2mm]
              & \quad + \partial^2 h^{\rho \sigma} \left( \partial_{\rho} \partial_{\sigma} h + \partial^2  h_{\rho \sigma} \right) - \frac12 \, \partial^{\rho} \partial^{\sigma} h \, \partial_{\rho} \partial_{\sigma} h, \notag \\[2mm]
   T_{\mu\nu} &= \overline{\chi}  \gamma_{(\mu} D_{\nu)} \chi \notag \\
   &= \overline{\chi}  \gamma_{(\mu} \partial_{\nu)} \chi + \frac12 h_{\rho (\mu} \overline{\chi}  \gamma^{\rho} \partial_{\nu)} \chi + \frac14 \partial_{\sigma} h_{\rho  (\nu} \overline{\chi}  \gamma_{\mu)} \gamma^{\rho \sigma} \chi + \frac18 h_{ \tau (\mu|} \partial_{\sigma} h_{\rho |\nu)} \overline{\chi}  \gamma^{\tau} \gamma^{\rho \sigma} \chi,
\end{align}
where on the r.h.s.\  all indices are lowered and raised with the Minkowski metric and $\gamma_{\mu}$ is the the flat gamma-matrix.

At second-order in the metric perturbation, the curvature-squared quantities can also be written in momentum space as:  
\begin{align}
 \textrm{Riem}^2 &= \int d^4 y  \,  d^4z \int\frac{d^4p}{(2 \pi)^4}  \frac{d^4q}{(2 \pi)^4} e^{-i p\cdot (x - y) - i q\cdot (x - z)} h^{\rho \sigma}(y)   h^{\a \b}(z) \notag \\[3mm] 
     & \qquad \times\Bigg\{   (p\cdot q)^2 \eta_{\a (\rho} \eta_{\sigma) \b} - 2\,  p\cdot q \, q_{(\rho} \eta_{\sigma) (\a}  p_{\b)} + \, q_{\rho} q_{\sigma} p_{\a} p_{\b} \Bigg\}, \label{riem2s} \\
 \textrm{Ric}^2 &= \int d^4 y  \,  d^4z \int\frac{d^4p}{(2 \pi)^4}  \frac{d^4q}{(2 \pi)^4} e^{-i p\cdot (x - y) - i q\cdot (x - z)} h^{\rho \sigma}(y)   h^{\a \b}(z) \notag \\[3mm]
     & \qquad \times\Bigg\{  \frac14 \, (p\cdot q)^2 \eta_{\rho \sigma} \eta_{\a \b}  +\frac14 \,  p^2\, q^2 \, \eta_{\a (\rho} \eta_{\sigma) \b} + \frac12 \, q^2 \,  \eta_{\rho \sigma} p_{\a} p_{\b} -  p \cdot q \, \eta_{\a \b} p_{(\rho} q_{\sigma)} \notag \\[3mm] & 
     \hspace{20mm} - \, q^2  \, p_{(\rho} \eta_{\sigma) (\a}  p_{\b)}  +  \frac12 \, p \cdot q  \,  p_{(\rho} \eta_{\sigma) (\a}  q_{\b)} +\frac12 \, p_{(\rho} q_{\sigma)} p_{(\a} q_{\b)}\Bigg\}, \label{ric2s} \\
R^2 &= \int d^4 y  \,  d^4z \int\frac{d^4p}{(2 \pi)^4}  \frac{d^4q}{(2 \pi)^4} e^{-i p\cdot (x - y) - i q\cdot (x - z)} h^{\rho \sigma}(y)   h^{\a \b}(z) \notag \\[3mm]
     & \qquad \times\Bigg\{  p^2\, q^2 \, \eta_{\rho \sigma} \eta_{\a \b} -2\, q^2\,  \eta_{\a \b} p_{\rho} p_{\sigma} +  \,p_{\rho} p_{\sigma} q_{\a} q_{\b} \Bigg\}, \label{r2s}\\
 \Box R &= \int d^4 y  \,  d^4z \int\frac{d^4p}{(2 \pi)^4}  \frac{d^4q}{(2 \pi)^4} e^{-i p\cdot (x - y) - i q\cdot (x - z)} h^{\rho \sigma}(y)   h^{\a \b}(z) \notag \\[3mm]
     & \qquad \times\Bigg\{  - \frac12 ( 2 \, (p\cdot q)^2 + p^2  \, p\cdot q )\, 
     \eta_{\rho \sigma} \eta_{\a \b} \notag \\[3mm] &  \hspace{15mm} 
     + \frac12 \, (2 \, (p^2)^2 + 3\, (p\cdot q)^2 + 7 \, p^2\, \, p\cdot q + 2 \, p^2 \,  q^2)) \, 
     \eta_{\a (\rho} \eta_{\sigma) \b}  \notag \\[3mm] &  \hspace{15mm}
+ \,(2\,  p^2 + 2 \,  p\cdot q + q^2) \,  \eta_{\rho \sigma} p_{\a} p_{\b} + \frac12 \, p\cdot q \, \eta_{\a \b} p_{\rho} p_{\sigma} \notag \\[3mm] &   \hspace{15mm}
+ ( p^2 +2 \, p \cdot q + 2 \, q^2) \, \eta_{\a \b} p_{(\rho} q_{\sigma)}  
  -2\, ( p^2 + 2\, p \cdot q +  q^2)  \, p_{(\rho} \eta_{\sigma) (\a}  p_{\b)}\notag \\[3mm] &  \hspace{15mm} 
  -2 \, ( p^2 + p \cdot q)  \,  p_{(\rho} \eta_{\sigma) (\a}  q_{\b)} - ( p^2 +  p\cdot q)  \, q_{(\rho} \eta_{\sigma) (\a}  p_{\b)}\notag \\[3mm]  &  \hspace{15mm}
 - p_{\rho} p_{\sigma} p_{\a} p_{\b} - p_{\rho} p_{\sigma} p_{(\a} q_{\b)}  \Bigg\}. \label{br2s}
\end{align}

\section{Boundary integrals} \label{app:int}
We here present some well-known results for scalar 2-point integrals, which are also referred
to as boundary integrals,
 \begin{align}
   &\int \frac{d^dk}{ \pi^{d/2}} \, \frac{k_{\mu}}{k^2 (p-k)^2} =   \frac{1}{2} p_{\mu} I(p), \label{2int1} \\
  &\int \frac{d^dk}{\pi^{d/2}} \, \frac{k_{\mu} k_{\nu}}{k^2 (p-k)^2} =  \frac{1}{4(d-1)} \left( d p_{\mu} p_{\nu} - p^2 \eta_{\mu \nu} \right) I(p),\\  
   &\int \frac{d^dk}{\pi^{d/2}} \, \frac{k_{\mu} k_{\nu} k_{\rho}}{k^2 (p-k)^2} =  \frac{1}{8(d-1)} \left((d+2) p_{\mu} p_{\nu} p_{\rho} -3 p^2 \eta_{(\mu \nu} p_{\rho)} \right) I(p),\\  
    &\int \frac{d^dk}{\pi^{d/2}} \, \frac{k_{\mu} k_{\nu} k_{\rho} k_{\sigma}}{k^2 (p-k)^2} =  \frac{1}{16(d^2-1)} \left( (d+4)(d+2) p_{\mu} p_{\nu} p_{\rho} p_{\sigma} \right.\notag \\
    & \hspace{60mm} \left. - 6 (d+ 2) p^2 \eta_{(\mu \nu} p_{\rho} p_{\sigma)} + 3 (p^2)^2 \eta_{(\mu \nu} \eta_{\rho \sigma)} \right) I(p),  \label{2int4}
 \end{align}
where $I(p) \equiv I(d\,|\,p) \equiv J(d;1,1,0)$, or more explicitly
\begin{align}
 I(p) &=   \int \frac{d^dk}{\pi^{d/2}} \, \frac{1}{k^2 (p-k)^2} \notag \\[3pt]
 &= - 2 i (d-1) (p^2)^{d/2-2} \,
i  \frac{\Gamma(1- \frac{d}{2}) \Gamma(\frac{d}{2})^2}{\Gamma(d)} = \frac{i}{\epsilon} + \left( 2 + \gamma - \log \frac{p^2}{\mu^2} \right)  + \mathcal{O}(\epsilon),\label{I(p)}
\end{align}
where $\gamma$ is the Euler-Mascheroni constant and $\mu$ is the renormalisation scale. 
Note the overall factor of $i$ which is a result of Wick rotating from Lorentzian signature.
The normalisation in the definition of $I(p)$ (and in (\ref{tensorint}) and (\ref{Jdnu}))
has thus been chosen in order to eliminate all factors of $\pi^{d/2}$ in the final expression.

\section{Useful formulae and identities} \label{app:id}

\begin{equation}
 \textrm{tr}\left(\gamma_{\mu} \gamma_{\nu} \gamma_{\rho} \gamma_{\sigma} \right) = \textrm{tr}\left(\gamma_{\mu} \gamma_{\sigma}\gamma_{\rho} \gamma_{\nu}  \right). \label{gamid}
\end{equation}

\begin{equation}
 \textrm{tr}\left( \gamma_{\mu} \gamma_{\nu} \gamma_{\rho}  \gamma_{\sigma} \gamma_{\a} \gamma_{\b} \right) =  \textrm{tr}\left( \gamma_{\rho} \gamma_{\nu} \gamma_{\mu}  \gamma_{\b} \gamma_{\a} \gamma_{\sigma} \right).
 \label{gamma6id}
\end{equation}

\begin{equation}
 \textrm{tr}\left( \gamma_{\mu} \gamma_{\nu} \gamma_{\rho}  \left(\gamma_{\sigma} \gamma_{\a} \gamma_{\b} + \gamma_{\b} \gamma_{\a} \gamma_{\sigma} \right) \right) = 2 \, \textrm{tr}\left( \gamma_{\mu} \gamma_{\nu} \gamma_{\rho}  \left(\eta_{\sigma \a} \gamma_{\b} - \eta_{\sigma \b} \gamma_{\a} + \eta_{\a \b} \gamma_{\sigma} \right) \right) .
 \label{gamma6id2}
\end{equation}

\begin{equation}
 \textrm{tr}\left( \gamma_{\mu} \gamma_{\nu} \gamma_{\rho} \left(\gamma^{\sigma} \gamma^{\a} \gamma^{\b} - \gamma^{\b} \gamma^{\a} \gamma^{\sigma} \right) \right) = - 12 \, \delta_{\mu \nu \rho}^{\sigma \a \b} \, \textrm{tr}\, {\bf 1}.
 \label{gamma6id3}
\end{equation}

Making use of the redefinition of the integration variable, letting $k \rightarrow -k + p$, and the gamma matrix identity \eqref{gamid}, it can be shown that
\begin{gather}
  \int \frac{d^dk}{(2 \pi)^d}  \textrm{tr}\biggl[ \frac{k\!\!\!/}{k^2}   \gamma_{\a}  
  \frac{(k\!\!\!/ - p\!\!\!/ )}{(k-p)^2} (2 k -p )_{(\rho} \gamma_{\sigma)} \biggr] =0. \label{niceid}
\end{gather}

Using identity \eqref{gamid}, 
 \begin{align}
&\int \frac{d^dk}{(2 \pi)^d} \, \frac{1}{k^2 (k- p)^2} \textrm{tr}\biggl\{( 2k\!\!\!/ - p\!\!\!/)  (2 k - p)_{\mu } \gamma_{\nu} ( 2k\!\!\!/ - p\!\!\!/)  (2 k -p )_{(\rho} \gamma_{\sigma)}\biggr\}, \notag \\[3pt]
  =& \, -32 \, i \,  T_{\mu \nu \rho \sigma}(p) + \int \frac{d^dk}{(2 \pi)^d} \, \frac{1}{k^2 (k- p)^2} \textrm{tr}\biggl\{ p\!\!\!/ \,  (2 k - p)_{\mu } \gamma_{\nu} \,  p\!\!\!/  (2 k -p )_{(\rho} \gamma_{\sigma)}\biggr\}. \label{niceid2}
\end{align}

Using identities \eqref{gamma6id} and \eqref{gamma6id2},
 \begin{align}
&\int \frac{d^dk}{(2 \pi)^d} \, \frac{1}{k^2 (k- p)^2} \textrm{tr}\biggl\{( 2k\!\!\!/ - p\!\!\!/)  (2 k - p)_{(\a } \gamma_{\b)} ( 2k\!\!\!/ - p\!\!\!/)  (2 k -p )_{\tau} \gamma_{\rho} \gamma_{\sigma} \gamma_{\nu} \biggr\}, \notag \\[2mm]
=& \, -32 \, i \left[ \eta_{\rho \sigma}  T_{\a \b \tau \nu}(p) - \eta_{\rho \nu}  T_{\a \b \tau \sigma}(p) +  \eta_{\sigma \nu}  T_{\a \b \tau \rho}(p) \right] \notag \\[2mm]
& \, + \int \frac{d^dk}{(2 \pi)^d} \, \frac{1}{k^2 (k- p)^2} \textrm{tr}\biggl\{p\!\!\!/  (2 k - p)_{(\a } \gamma_{\b)}  p\!\!\!/  (2 k -p )_{\tau} \gamma_{\rho} \gamma_{\sigma} \gamma_{\nu} \biggr\},\label{niceid3}
\end{align}
where we have also used equation \eqref{niceid2}.

Furthermore, using identities \eqref{gamma6id} and \eqref{gamma6id2} and 
\begin{equation}
 \textrm{tr} \left[ ( 2k\!\!\!/ - p\!\!\!/)  \gamma_{\nu} ( 2k\!\!\!/ - p\!\!\!/)  p\!\!\!/\right] =   \textrm{tr} \left[ p\!\!\!/  \gamma_{\nu} p\!\!\!/ p\!\!\!/\right] + 2(k^2- (k-p)^2) \textrm{tr} \left[ \gamma_{\nu} ( 2k\!\!\!/ - p\!\!\!/)  \right]- 2 (k^2+ (k-p)^2) \textrm{tr} \left[ \gamma_{\nu} p\!\!\!/  \right],
\end{equation}
we also have the following identity:
 \begin{align}
&\int \frac{d^dk}{(2 \pi)^d} \, \frac{1}{k^2 (k- p)^2} \textrm{tr}\biggl\{( 2k\!\!\!/ - p\!\!\!/)  \gamma_{\nu} ( 2k\!\!\!/ - p\!\!\!/)  (2 k - p)_{(\a } \gamma_{\b)} p\!\!\!/ (2 k - p)_{(\rho } \gamma_{\sigma)}  \biggr\}, \notag \\[2mm]
=& \, -32 \, i \left[ p_{(\a} T_{\b) \nu \rho \sigma}(p) - p_{(\rho}  T_{  \sigma) \nu \a \b}(p) \right] \notag \\[2mm]
& + \,\int \frac{d^dk}{(2 \pi)^d} \, \frac{1}{k^2 (k- p)^2} \textrm{tr}\biggl\{ p\!\!\!/  \gamma_{\nu} p\!\!\!/  (2 k - p)_{(\a } \gamma_{\b)} p\!\!\!/ (2 k - p)_{(\rho } \gamma_{\sigma)}  \biggr\},\label{niceid4}
\end{align}

 \section{Symmetrisation of the two-point function} \label{app:subanti2pt}
 We show that in the integral expression for the two-point function, equation \eqref{2pt}, the antisymmetrised part of the integral in indices $\mu \nu$ vanishes,  
 \begin{align}
\int \frac{d^d k}{(2 \pi)^d} \textrm{tr} \left\{ \frac{1}{k^2(k-p)^2} \left( k\!\!\!/ (2 k - p)_{[\mu} \gamma_{\nu]} (k\!\!\!/ - p\!\!\!/) (2 k - p)_{(\rho} \gamma_{\sigma)} \right) \right\} = 0. \label{anti2pt}
\end{align}
First we observe that under the following reparametrisation of the integration variable, $k \rightarrow -k +p,$ 
\begin{equation}
 \frac{1}{k^2(k-p)^2} \rightarrow \frac{1}{k^2(k-p)^2}, \quad (2 k - p) \rightarrow - (2 k - p). \label{transf}
\end{equation}
Hence we rewrite the integral in equation \eqref{anti2pt} as
 \begin{align}
&\frac14 \int \frac{d^d k}{(2 \pi)^d} \textrm{tr} \left\{\frac{1}{k^2(k-p)^2} \left((2 k\!\!\!/ - p\!\!\!/)  (2 k - p)_{[\mu} \gamma_{\nu]} (2 k\!\!\!/ - p\!\!\!/) (2 k - p)_{(\rho} \gamma_{\sigma)}  \right) \right\}\notag \\[3pt]
-& \frac14 \int \frac{d^d k}{(2 \pi)^d} \textrm{tr}\left\{ \frac{1}{k^2(k-p)^2} \left( p\!\!\!/  (2 k - p)_{[\mu} \gamma_{\nu]}  p\!\!\!/ (2 k - p)_{(\rho} \gamma_{\sigma)}  \right)\right\}.
\end{align}
Now we note that, by symmetry properties of the trace, the above integrals are symmetric under the simultaneous exchange $\mu \leftrightarrow \rho$ and $\nu \leftrightarrow \sigma,$ but the only potential non-zero term that we can write down for these integrals is
\begin{equation*}
 p_{[\mu} \eta_{\nu] (\rho} p_{\sigma)}, 
\end{equation*}
which is antisymmetric under the aforementioned exchange. Hence equation \eqref{anti2pt} is established. 

\section{Evaluation of $\tilde{T}$} \label{app:Ttilde}

In this appendix we evaluate $\tilde{T}_{\nu \rho \sigma \a \b}(p,q),$ defined in equation \eqref{Ttildedef},
 \begin{align}
i \int \frac{d^dk}{(2 \pi)^d} \, \textrm{tr}\biggl\{ \frac{k\!\!\!/ }{k^2} \left( \gamma_{\nu} (k\!\!\!/ + p\!\!\!/ ) + (2 k + p)_{\nu} \right)  (2 k + p - q )_{(\rho} \gamma_{\sigma)} \frac{k\!\!\!/ - q\!\!\!/ }{(k-q)^2}  (2k - q )_{(\a} \gamma_{\b)} \biggr\}. \label{int1}
\end{align}
We first note that each factor can be written as a sum of a term proportional to $(2k-q)$ and another $k$-independent term. For example,
\begin{equation}
 k_{\mu} = \frac12 (2k-q)_{\mu} + \frac12 q_{\mu}.
\end{equation}
Now, again using the trick in appendix \eqref{app:subanti2pt} that under a reparametrisation of the integration variable $ k \rightarrow - k + q$ we have 
\begin{equation}
 \frac{1}{k^2(k-q)^2} \rightarrow \frac{1}{k^2(k-q)^2}, \quad (2 k - q) \rightarrow - (2 k - q),
\end{equation}
we rewrite integral \eqref{int1} in terms of an even number of $(2k-q)$ factors, 
 \begin{align}
\frac{i}{8} \int \frac{d^dk}{(2 \pi)^d} \, \frac{1}{k^2 (k-q)^2}\textrm{tr}\biggl\{ & (2 k\!\!\!/ - q \!\!\!/ ) \, \gamma_{\nu}  \,(2 k\!\!\!/ - q \!\!\!/ )  \,(2 k - q )_{(\a} \gamma_{\b)} \, (2 k\!\!\!/ - q \!\!\!/ )  \, p_{(\rho} \gamma_{\sigma)} \notag \\[2pt]
 - & (2 k\!\!\!/ - q \!\!\!/ ) \, \gamma_{\nu}  \,(2 k\!\!\!/ - q \!\!\!/ )  \, (2 k - q )_{(\a} \gamma_{\b)}  \, q \!\!\!/   \, (2k - q )_{(\rho} \gamma_{\sigma)} \notag \\[2pt]
+ & (2 k\!\!\!/ - q \!\!\!/ )  \,\gamma_{\nu} \,  q \!\!\!/  \, (2 k - q )_{(\a} \gamma_{\b)}  \, (2 k\!\!\!/ - q \!\!\!/ )  \,  (2k - q )_{(\rho} \gamma_{\sigma)} \notag \\[2pt]
+ & (2 p\!\!\!/ + q \!\!\!/ )  \,\gamma_{\nu} \, (2 k\!\!\!/ - q \!\!\!/ )  \, (2 k - q )_{(\a} \gamma_{\b)}  \,  (2 k\!\!\!/ - q \!\!\!/ )  \,  (2k - q )_{(\rho} \gamma_{\sigma)} \notag \\[2pt]
- & (2 k\!\!\!/ - q \!\!\!/ )  \,\gamma_{\nu}  \, q \!\!\!/  \, (2 k - q )_{(\a} \gamma_{\b)}  \,  q \!\!\!/  \, p_{(\rho} \gamma_{\sigma)} \notag \\[2pt]
- & (2 p\!\!\!/ + q \!\!\!/ )  \,\gamma_{\nu}  \,(2 k\!\!\!/ - q \!\!\!/ )  \, (2 k - q )_{(\a} \gamma_{\b)} \,  q \!\!\!/   \, p_{(\rho} \gamma_{\sigma)} \notag \\[2pt]
+ & (2 p\!\!\!/ + q \!\!\!/ )  \,\gamma_{\nu}  \, q \!\!\!/  \,  (2 k - q )_{(\a} \gamma_{\b)} \, (2 k\!\!\!/ - q \!\!\!/ )  \, p_{(\rho} \gamma_{\sigma)} \notag \\[2pt]
- & (2 p\!\!\!/ + q \!\!\!/ )  \,\gamma_{\nu} \,  q \!\!\!/  \, (2 k - q )_{(\a} \gamma_{\b)} \, (2 k\!\!\!/ - q \!\!\!/ )  \,  (2k - q )_{(\rho} \gamma_{\sigma)} \notag \\[2pt]
+ & 2\, (2 k - q )_{\nu} \, (2 k\!\!\!/ - q \!\!\!/ )  \,(2 k - q )_{(\a} \gamma_{\b)} \, (2 k\!\!\!/ - q \!\!\!/ )  \, p_{(\rho} \gamma_{\sigma)} \notag \\[2pt]
+ & 2\, (q -p )_{\nu}   \,(2 k\!\!\!/ - q \!\!\!/ )  \, (2 k - q )_{(\a} \gamma_{\b)} \,(2 k\!\!\!/ - q \!\!\!/ )   \, (2k - q )_{(\rho} \gamma_{\sigma)} \notag \\[2pt]
- & 2\, (2 k - q )_{\nu} q \!\!\!/ \,(2 k - q )_{(\a} \gamma_{\b)} \,q \!\!\!/ \, p_{(\rho} \gamma_{\sigma)} \notag \\[2pt]
- & 2\, (q -p )_{\nu}   \,q \!\!\!/ \, (2 k - q )_{(\a} \gamma_{\b)} \,q \!\!\!/ \, (2k - q )_{(\rho} \gamma_{\sigma)} \biggr\},
\end{align}
cancelling some terms using the gamma matrix identity \eqref{gamid}.
Using identity \eqref{gamma6id3}, we can show that the terms on the sixth and seventh line in the expression above cancel. Furthermore, we can use identities \eqref{niceid2}, \eqref{niceid3} and \eqref{niceid4} to simplify the expressions on the ninth and tenth lines; first, third and fourth lines and the second line of above expression, respectively. Collecting all the terms we find that 
 \begin{align}
\tilde{T}_{\nu \rho \sigma \a \b}(p,q) = \, 8 \, \left[ 3 \, p_{(\rho}\,  T_{\sigma) \nu \a \b}(q) + 2 \,  (p+ q)_{\nu} \, T_{ \rho \sigma \a \b}(q) - p^{\tau} \, \eta_{\nu (\rho} \,  T_{\sigma) \tau \a \b}(q) \right]. \label{int1ans}
\end{align}

\end{document}